# Employing magma ocean crystallization models to constrain structure and composition of the lunar interior

Sabrina Schwinger[a] and Doris Breuer[a]

[a]German Aerospace Center (DLR), Rutherfordstr. 2, 12489 Berlin, Germany. sabrina.schwinger@dlr.de, doris.breuer@dlr.de

**Correspondence to:** Sabrina Schwinger, sabrina.schwinger@dlr.de

**Abstract**

The process of lunar magma ocean solidification provides crucial constraints on the composition and extent of distinct chemical reservoirs in the lunar mantle that formed during the early evolution of the Moon. We use a combination of phase equilibria models consistent with recent experimental results on fractional crystallization of the lunar magma ocean to study the effect of bulk silicate Moon composition on the properties of lunar mantle reservoirs.

We find that the densities and relative proportions of these mantle reservoirs, in particular of the late forming ilmenite bearing cumulates (IBC), strongly depend on the FeO content of the bulk silicate Moon. This relation has implications for post-magma ocean mantle dynamics and the resulting mass distribution in the lunar interior, because the IBC form at shallow depths but tend to sink towards the core mantle boundary due to their high density. We quantify the relations between bulk silicate Moon FeO content, IBC thickness and bulk Moon density as well as mantle stratigraphy and bulk silicate Moon moment of inertia in order to constrain the bulk silicate Moon FeO content and the efficiency of IBC sinking.

In combination with seismic and selenodetic constraints on mantle stratigraphy, core radius and extent of the possibly IBC bearing low velocity zone at the core mantle boundary as well as considerations about the present day selenotherm and the effects of reservoir mixing by convection our model indicates that the bulk silicate Moon is only moderately enriched in FeO compared to the Earth's mantle and contains most likely about 9.4 – 10.9 wt% FeO (with a lowermost limit of 8.3 wt % and an uppermost limit of 11.9 wt%). We further conclude that the observed bulk silicate Moon moment of inertia requires that sinking of the IBC layer by mantle convection was incomplete: only ~ 20 – 60 % of the IBC material might have reached the core mantle boundary, while the rest either remained at the depth of its formation right beneath the crust or was mixed into the middle mantle.

# 1. Introduction

Estimates of the bulk silicate Moon (BSM) composition have been suggested based on a number of different geochemical, petrological and geophysical arguments but have yet to arrive at a general consensus. Analyses of lunar rocks indicate that refractory lithophile trace element ratios in the lunar mantle are chondritic (e.g. Wänke, 1981; Jochum, 1986a,b), which suggests that also refractory main elements like Ca and Al occur in chondritic ratios in the BSM (e.g. Jagoutz et al., 1979; Hart and Zindler, 1986). However, additional constraints on the BSM composition are difficult to obtain using petrological arguments, since direct samples from the lunar mantle are not available.

Materials derived from partial mantle melting, like pristine glasses and mare basalts have been used to infer the compositions and depths of their mantle sources (e.g. Papike et al., 1976; Binder, 1982, 1985; Beard et al., 1998; Hallis et al., 2014) and derive a range of possible BSM compositions (e.g. Jones and Delano, 1989; Warren, 2005), but also indicate that the mantle is compositionally heterogeneous, which makes it difficult to directly infer BSM properties. Seismic data confirm that the lunar mantle must be chemically stratified, since the seismic velocities cannot be fitted by a single compositional model (e.g. Kuskov and Fabrichnaya, 1994).

With a heterogeneous lunar mantle, estimates of the BSM composition require additional information about the relative proportions of individual mantle reservoirs, which can be inferred by reconstructing the Moon's formation history. The Moon is thought to have experienced an early stage of planetary scale melting that lead to the formation of a global magmasphere, which is commonly referred to as the lunar magma ocean (e.g. Smith et al., 1970; Wood et al., 1970; Warren, 1986; Rai and van Westrenen, 2014; Steenstra et al., 2016). The fractional crystallization of this lunar magma ocean (LMO) resulted in a stratified lunar mantle and the formation of the Moon's anorthositic crust. If the different chemical reservoirs composing the lunar mantle and crust indeed formed from a global lunar magma ocean, their compositions and proportions are not independent but linked through the process of fractional crystallization from a common parent magma. These relations can be used in combination with compositional data from lunar samples to estimate the BSM composition (e.g. Jones and Delano, 1989). However, the resulting BSM compositions are sensitive to assumptions made about the degree of crystal fractionation in the magma ocean. Most notably, the FeO content of the BSM varies substantially depending on the assumed degree of fractionation (Jones and Delano, 1989), making it one of the least well constrained aspects of the bulk silicate Moon composition. Warren (2005) argued that the MgO/FeO ratio of the lunar magma ocean needs to be sufficiently high to be consistent with the formation of Mg-rich olivines with $Fo_{87-94}$ that have been observed in Mg-suite rocks and melt breccias. However, recent models of LMO cumulate convection have shown that early olivine cumulates can partially melt during convective ascent (Maurice et al., 2020), which would lead to further depletion of the cumulate in Fe. This process could produce Mg-rich olivine even if the LMO MgO/FeO is too low to crystallize Mg-rich olivine directly from the melt, which relaxes the constraints on the required BSM FeO content.

Since Fe is the heaviest of the main elements comprising the Moon, the BSM FeO content is closely related to the BSM density, which can be inferred from measurements of the bulk Moon density (e.g. Matsumoto et al., 2015) and estimates of the size and density of the lunar core (e.g. Garcia et al., 2011; Weber et al., 2011; Antonangeli et al., 2015). Due to the close relation of FeO content and rock density, the distribution of FeO in the lunar mantle is likely to affect the BSM moment of inertia, which is known from selenodetic data (e.g. Matsumoto et al., 2015). The distribution of FeO in the lunar mantle also affects its seismic properties, but gradual changes in FeO contents throughout the mantle are difficult to resolve (Gagnepain-Beyneix et al., 2006) and have hence not been explicitly considered in seismic models yet.

The possible distributions of chemical reservoirs in the lunar interior can be estimated by modeling not only the formation of the reservoirs by LMO crystallization but also their subsequent relocation or mixing by solid state convection. Since FeO progressively accumulates in the LMO during crystallization, the resulting cumulate stratigraphy is gravitationally unstable, with denser FeO-rich material overlying lighter FeO-poor cumulate rocks (e.g. Hess and Parmentier, 1995; Elkins-Tanton et al., 2011). This generally enables an overturn of the primary stratigraphy, that can be simulated by solid state convection models (e.g. Yu et al., 2019). However, the efficiency of this overturn process is still poorly constrained due to large uncertainties in the rheological properties of individual mantle reservoirs.

To investigate the relation between the BSM FeO content and the physical properties of the chemical reservoirs formed by LMO solidification, we a) develop a LMO crystallization model that is consistent with recent fractional crystallization experiments, b) set up a simple model to simulate different mantle overturn scenarios by varying mantle stratigraphies and degrees of mixing and c) determine the mass distribution in the lunar interior for different stratigraphic models to test their consistency with the bulk Moon density and BSM moment of inertia.

In the following we will describe our model of LMO solidification (section 2), present the effects of BSM FeO content on the properties of the modeled LMO cumulate (section 3), describe our lunar interior structure models (section 4), present the physical properties of our lunar interior models and their consistency with the observed physical properties of the Moon (section 5) and discuss the applicability and limitations of our models for constraining BSM FeO contents.

## 2. Modeling of LMO solidification

In the following we outline the details of our LMO solidification model and discuss a) the properties of the LMO such as depth and composition and its crystallization process and b) the applicability of different magma crystallization softwares to the LMO scenario considering recent fractional crystallization experiments.

### 2.1. Lunar magma ocean properties

In our LMO model we assume that the initial magma ocean comprised the entire silicate Moon. We further assume that the initial LMO had a composition similar to the Earth's mantle with possible FeO contents between 8 and 13 wt%, and that its solidification occurred by pure fractional crystallization. The reasoning behind these assumptions and the chosen parameter ranges is detailed below.

#### *2.1.1. LMO depth*

The Moon is thought to have formed from the debris produced by a giant collision of the proto-Earth with a smaller protoplanet, However, the initial thermal state of the fully accreted Moon and the initial depth of a putative lunar magma ocean (LMO) still remain unclear although various scenarios have been addressed in computational studies in the recent years (Asphaug et al., 2014; Canup et al., 2015, Lock and Stewart, 2017; Nakajima and Stevenson, 2018).

One important constraint on the minimum magma ocean depth is the thickness of the anorthositic crust. The amount of crystallizing plagioclase and thus the crustal thickness is a function of magma ocean composition (i.e. $Al_2O_3$ and CaO content), magma ocean depth and the efficiency of plagioclase floatation. By choosing a magma ocean composition and assuming that all crystallizing plagioclase is incorporated into the crust, it is possible to constrain the minimum LMO depth required to produce the observed thickness of the lunar crust.

It has been argued that the crustal thickness could also be used to constrain an upper limit of magma ocean depth to ~ 600 – 800 km (Elkins-Tanton et al., 2011; Charlier et al., 2018), relying on the assumption that all plagioclase that is able to buoyantly float in a magma ocean will be incorporated in the anorthositic crust. Indeed, the compositions of high-Ti mare basalts suggest that their mantle sources contain only very limited amounts of plagioclase. However, the compositions of anorthositic and troctolithic cumulates in terrestrial magma chambers (e.g. Namur et al., 2011) indicate that in such a setting only a limited fraction of positively buoyant plagioclase actually floats and forms anorthosite, while a larger fraction is incorporated in the bottom cumulate of the magma chamber. Even if the crystallization behavior of a terrestrial magma chamber cannot be readily applied to a global magma ocean, it illustrates the fact that it is not safe to assume perfect efficiency of plagioclase floatation (e.g. Snyder et al., 1992). Although there is no direct evidence for a plagioclase-rich reservoir in the lunar mantle, it is possible that such a reservoir has either not been sampled by known igneous lunar rocks or the plagioclase-bearing cumulates were dense enough to sink to sufficient depths to enter the stability field of garnet. In addition, the crust thickness might have been limited by the presence of water in the lunar magma ocean, as plagioclase crystallization is impeded in a water-bearing magma (e.g. Lin et al., 2017, 2020). For these reasons, the crust thickness alone can only be used to determine lower limits of magma ocean depth or $Al_2O_3$ content.

Another possible constraint on the LMO depth is the depth of the mantle source regions of green picritic glasses (Longhi et al., 2006). Any primitive lower mantle of bulk silicate Moon composition would have too high $Al_2O_3$ contents to qualify as a source region for the $Al_2O_3$-poor glasses. Therefore, the glass source regions, which are supposed to lie at a depth of 700-1000 km (Longhi et al., 2006), must consist of $Al_2O_3$-poor magma ocean cumulates. Such cumulates must have crystallized early and were located at the base of the MO. Considering possible mixing of the cumulate pile by mantle convection and the depths of the source regions, it seems likely that the magma ocean was at least as deep as the mantle sources of the green picritic glasses: since the

Al$_2$O$_3$-poor cumulates are less dense than later cumulates or a primitive lower mantle, they would rise and not sink during mantle overturn.

Furthermore, metal–silicate partitioning models by Rai and van Westrenen (2014) suggest that metal–silicate equilibrium during lunar core formation occurred at depths close to the present-day lunar core–mantle boundary, which is consistent with a deep magma ocean involving the whole BSM. Based on this evidence and the arguments above we assume that the bulk silicate Moon was completely molten, forming a deep lunar magma ocean.

*2.1.2. LMO composition*

The compositions of LMO and BSM are identical if the lunar magma ocean included the whole bulk silicate Moon. In the case of a shallow LMO, the BSM comprises both magma ocean cumulates and a primitive mantle. The possible composition of the LMO and the primitive mantle then depend on how the primitive mantle was affected by melting. If the primitive mantle never experienced any melting, then both the primitive mantle and the LMO would have BSM compositions. However, such a scenario would require a sharp thermal boundary between LMO and primitive mantle, so that it seems more likely that a primitive mantle was at least partially molten. In this case, at least part of the primitive mantle would be depleted in incompatible elements, while the LMO would be complementarily enriched in incompatible elements. As detailed above, we assume that the LMO included the whole bulk silicate Moon, so that in the further course of this paper LMO composition and BSM composition are considered identical.

Though an enrichment of the bulk silicate Moon in Al and Ca compared to the Earth's mantle has been suggested in earlier studies (e.g. Taylor and Jakes, 1974; Taylor, 1982, commonly referred to as Taylor whole Moon (TWM) model), later observations like the isotopic similarity of Earth and Moon (Wiechert et al., 2001; Touboul et al., 2007; Zhang et al., 2012; Dauphas et al., 2014) and refined, lower estimates of anorthositic crust thicknesses (Taylor et al., 2013; Wieczorek et al., 2013) lead to the consensus that the composition of the bulk silicate Moon is probably largely Earth-mantle-like (e.g. Hauri et al., 2015; O'Neill, 1991).

An exception to that is the FeO content of the bulk silicate Moon, which has been suggested to be significantly higher than in the Earth's mantle, based on both petrological (e.g. Ringwood, 1979; Delano and Lindsley, 1983) and geophysical arguments (e.g. Buck and Toksöz, 1980; Nakamura, 1983). However, there is no clear consensus on the degree of Fe enrichment and suggested BSM FeO contents vary significantly between ~8 – 17 wt% (e.g. Jones and Delano, 1989; Warren, 1986 and references therein), though most studies seem to favor a moderate BSM FeO content of ~12 – 13 wt% (e.g. Elkins-Tanton et al., 2011 and references therein).

Consequently, we decided to assume BSM compositions based on the BSM composition proposed by O'Neill (1991), which is similar to the Earth's mantle except for a moderate enrichment in FeO. To study the effects of varying Fe/Mg ratios, we varied BSM FeO contents and adjusted the respective MgO contents accordingly. LMO fractional crystallization experiments are available for LMO FeO contents of about 8 – 13 wt% (Rapp and Draper, 2018; Charlier et al., 2018), so we chose this range of compositions to make sure that our LMO fractional crystallization models can be directly compared to experimental results.

### 2.1.3. Degree of crystal fractionation

The efficiency of crystal fractionation during magma ocean crystallization influences the degree by which the solid cumulates can chemically equilibrate with the remaining melt. This affects the distribution of FeO in the magma ocean cumulate. Since Fe is an incompatible element, Fe accumulates in the remaining melt as the LMO solidifies. In an equilibrium crystallization scenario, early cumulates can equilibrate with the remaining melt, allowing them to progressively enrich their initially Fe-poor composition in Fe. During fractional crystallization the cumulate is physically separated from the melt, so that cumulate cannot equilibrate with later more evolved melt compositions. In a cumulate formed by equilibrium crystallization Fe is hence more evenly distributed. In a cumulate formed by fractional crystallization Fe is more strongly concentrated in the late cumulates, while the early cumulates are depleted in Fe. The degree of crystal fractionation in a magma ocean depends on several parameters, including the dominant location of crystal nucleation and growth (i.e. in cold downwellings in the magma or on the magma ocean floor) and the efficiency of crystal entrainment by the convecting magma, that are poorly constrained, so that to date the degree of fractionation during LMO crystallization remains unknown.

In our model, we assume pure fractional crystallization. This assumption tends to maximize the compositional differences among mantle reservoirs and hence has to be seen as an extreme case of compositional differentiation. However, as described below, we also consider mixing and compositional homogenization of the primary mantle reservoirs. Equilibrium crystallization would produce degrees of chemical inhomogeneity that are between both extreme cases. Hence, we make sure to cover the whole range of possible degrees of chemical inhomogeneity in the lunar mantle by assuming pure fractional crystallization in our LMO solidification model.

## 2.2. LMO solidification models

Several attempts have been made to estimate the mineralogical composition of the magma ocean cumulates based on theoretical calculations of phase relations (e.g. Snyder, 1992; Elkins-Tanton et al., 2011; Warren and Wasson, 1979, Longhi, 2003). However, as it has been noted by Elkins-Tanton et al. (2011), experiments covering relevant compositions and physical conditions are required to confirm the results of theoretical calculations and common assumptions of cumulate mineralogies.

Within the last decade several experimental studies have addressed lunar magma ocean crystallization, adopting different starting compositions and conditions of crystallization. Elardo et al. (2011) studied equilibrium crystallization of Ca-Al-enriched "Taylor Whole Moon" (TWM, e.g. Taylor, 1982) and Earth mantle-like "Lunar Primitive Upper Mantle" (LPUM, e.g. Longhi 2003, 2006; Warren, 2005) compositions at different pressures and temperatures ranging from 0.5-4 GPa and 1050-1825 °C. They found that olivine is the liquidus phase for all relevant LMO compositions and pressure ranges in the early phase of magma ocean crystallization. Lin et al. (2017, 2020) performed experiments simulating fractional crystallization of a 700 km deep magma ocean

(corresponding to crystallization pressures of 0 – 3 GPa) with LPUM composition and various water contents, demonstrating that the water content influences the plagioclase/clinopyroxene ratio that crystallizes from the melt, as it has been shown in earlier studies (e.g. Sisson and Grove, 1993). Charlier et al. (2018) simulated fractional crystallization of shallow magma oceans of 500 – 600 km depth by subtracting a theoretical olivine component from 7 different initial magma ocean compositions and by performing fractional crystallization experiments with the remaining compositions. Rapp and Draper (2018) carried out stepwise fractional crystallization experiments with a LPUM composition and a pressure range of 0 – 4 GPa, assuming an initial magma ocean depth of about 1400 km. As noted by Rapp and Draper (2018), their results are not consistent with the predictions of the frequently adopted petrological model by Snyder et al. (1992), which implies the necessity of re-evaluating currently used models and explicitly test their applicability to a LMO crystallization scenario.

Snyder et al. (1992) used fractional and equilibrium crystallization programs, which were developed specifically for modeling lunar magma ocean solidification (Longhi 1980, 1982). These programs calculate phase abundances and compositions based on a set of algorithms and experimentally determined liquidus boundaries in various subprojections in the olivine-plagioclase-wollastonite-silica system. These algorithms have been developed further over time and applied to magma crystallization scenarios on various planetary bodies (e.g. Neal et al., 1994; Brown and Elkins-Tanton, 2011; Thompson et al., 2003; Slater et al., 2003; Lin et al., 2017, 2020). The most recent versions of the crystallization algorithms are available in SPICEs, a Matlab environment developed by Davenport et al. (2013).

Other studies also used the MELTS and pMELTS algorithms to model LMO solidification (e.g. Arai and Maruyama, 2017). The MELTS and pMELTS algorithms calculate phase properties by Gibbs energy minimization using a database of experimentally determined thermodynamic properties of minerals and silicate melt. pMELTS is specifically calibrated for peridotite compositions at elevated pressures (1-3 GPa), while MELTS covers a larger range of compositions and is most reliable at low pressures (0-3 GPa). MELTS and pMELTS have been widely used to study terrestrial magmatic systems but has also been applied to study extraterrestrial magmatic rocks (e.g. Slater et al., 2003; Thomson et al., 2003).

*2.2.1. Reproduction of experiments by LMO crystallization models*

In order to test the consistency of commonly used magma solidification models with experimental results, we fitted the results of crystallization experiments by Rapp and Draper (2018) with FXMOTR, MELTS and pMELTS. Thereby we attempted to reproduce both individual crystallization experiments, assuming the conditions used by Rapp and Draper (2018), and the complete crystallization sequence, starting from the initial pressure and composition used by Rapp and Draper (2018) in their first crystallization step. In addition, we tested different combinations of crystallization algorithms in order to find a model that fits best the experimental data in terms of mineral modal abundances and degrees of solidification for given pressures, temperatures and compositions. The results of all calculations and fits to experimental data are given in the supplementary material.

As it has been noted by Ghiorso et al. (2002), the pMELTS algorithm overestimates the stability of garnet at high pressures. According to experimental studies, e.g. by Elardo et al. (2011), garnet is not a liquidus phase in the lunar magma ocean and is hence unlikely to form in a fractionally crystallizing magma ocean. Therefore, we suppressed garnet crystallization in the MELTS/pMELTS crystallization model. FXMOTR stipulates that olivine is the liquidus phase at the beginning of magma ocean crystallization, so that early garnet crystallization is excluded by default.

We find that neither pMELTS/MELTS nor FXMOTR were successful in reproducing all aspects of the compositional evolution of the liquid (see Fig. 1) and the mineral modal abundances in the cumulate (see supplementary material) due to their specific limitations. The main problem of the pMELTS algorithm in predicting the correct mineralogies lies in the underestimation of olivine stability at high pressures (Ghiorso et al. 2002), that leads to crystallization of orthopyroxene before olivine in deep LMO settings. FXMOTR on the other hand correctly predicts the transition from olivine to orthopyroxene crystallization in the early stages of LMO solidification but underestimates later liquid $Al_2O_3$ and $TiO_2$ contents and hence overestimates the amounts of crystallizing plagioclase and ilmenite.

However, both models complement each other in that FXMOTR succeeded in reproducing the early crystallization history, while MELTS/pMELTS produced accurate results in the late crystallization stages. Therefore, we tested a combined modeling approach where FXMOTR was used for early and MELTS for later steps of fractional crystallization. Indeed, the best fit of the crystallization sequence and the chemical evolution of the remaining liquid was achieved by a combined model in which the early stages of crystallization (up to ~ 45 pcs, just before Opx becomes stable) are calculated with FXMOTR and the late stages are modeled with pMELTS/MELTS (Fig. 1). This modeling approach also produces good fits for relative mineral abundances and crystallization temperatures (see supplementary material).

To test the applicability of this modeling approach to other compositions and magma ocean depths, we fitted experimental results by Charlier et al. (2018) for an FeO-rich LMO composition proposed by O'Neill (1991), with a low MgO/(MgO+FeO) ratio of only 0.74 compared to 0.81 for the LPUM composition used by Rapp and Draper (2018). The evolution of the cumulate mineralogies during crystallization as well as the crystallization temperatures are well reproduced by our crystallization model. These results indicate that our modeling approach is generally applicable to magma oceans with variable #Mg, terrestrial refractory element contents and depths of ~ 600 – 1400 km.

## 3. Effect of FeO content on LMO cumulate reservoirs

The results of our lunar magma ocean crystallization models indicate that changing the FeO content of the bulk LMO affects the compositions, volumes and densities of individual mantle reservoirs. To enable a systematic quantification of these changes, we simplified the complex compositional layering resulting from the fractional crystallization models to a few compositional reservoirs by grouping adjacent compositional layers of similar mineralogy and density. The resulting reservoirs are an olivine-dominated lower mantle (LM), a pyroxene-dominated upper mantle (UM), Ti-rich, high density material forming in the late stages of magma ocean solidification (IBC), cumulates forming from the melt remaining after IBC solidification (KREEP) and a crust that consists of a mixture of plagioclase and upper mantle material with their relative amounts chosen as to reach a crustal density of 2900 kg/m³ (Huang and Wieczorek, 2012). The crust volume was chosen as to fit a crustal thickness of 40 km, which is consistent with estimates by Wieczorek et al. (2013). Any excess plagioclase formed from the magma ocean was assumed to be trapped in the upper mantle (UM).

Increasing the BSM FeO content generally leads to increasing densities of individual compositional reservoirs and influences the onset of IBC crystallization and hence the proportions of the IBC and UM reservoirs. In models with higher FeO content, high density phases like Fe-Ti-Oxides or fayalite crystallize earlier in the lunar magma ocean, while in low FeO models they appear later in the crystallization sequence. In low FeO models the formation of dense IBC material starts with the formation of Fe-Ti-Oxides, which is mainly triggered by the oversaturation of the melt in $TiO_2$. In this case the main silicate species in the IBC material is clinopyroxene. In high FeO models the high concentrations of FeO lead to the formation of dense fayalite before Fe-Ti-Oxides start to form, so that the IBC material contains olivine as the main silicate species and lower amounts of clinopyroxene. As a consequence, the FeO content correlates directly with the thickness of the IBC layer, i.e. the higher the FeO content the thicker the IBC layer. For BSM compositions with 8 – 13 wt.% FeO, the thickness increases from about 11.6 km to 31.1 km, respectively (Fig. 2).

As Fe accumulates in the LMO melt during fractional crystallization, the earlier cumulates experience a lower enrichment in FeO with increasing BSM FeO content compared to the late cumulates. An increase in BSM FeO content by 4wt% leads to an increase of LM and UM FeO contents by only ~3wt%, but an increase of IBC FeO contents by ~6.4 wt%. Accordingly, the density changes are slightly smaller (~45 kg/m³) for the LM and UM reservoirs and higher (~55 kg/m³) for the IBC reservoir for a 4wt% increase in BSM FeO content.

## 4. Lunar interior structure models

The magma ocean crystallization model determines the densities and mineralogical compositions of different chemical reservoirs in the lunar interior, assuming the specific pressure and temperature conditions of their formation during bottom up crystallization of the LMO. However, the pressure and temperature conditions for each individual reservoir changed during further evolution of the lunar interior, as a) the Moon progressively cooled down to today's selenotherm and b) solid state convection changed the spacial distribution of the reservoirs. In addition, solid state convection might have led to mixing and chemical equilibration of the primary cumulate layers. All these processes led to changes in the density of the chemical reservoirs in the lunar interior, which need to be quantified in order to obtain lunar interior models that can be compared directly to the present-day physical properties of the Moon.

To this end, we set up a set of simple models accounting for the possible effects of mixing, changes in the stratigraphy, and interior cooling since the crystallization of the primary chemical reservoirs. Based on the compositions and mass fractions of the primary chemical reservoirs formed by LMO solidification, we determined the compositions of mixed reservoirs that would result from merging two or more primary reservoirs. For each chemical reservoir we calculated the material density as a function of depth, considering different possible selenotherms. These density functions allow us to determine the mass distribution in different lunar interior models, which consider the possible arrangement of the BSM chemical reservoirs in different stratigraphic configurations as well as a range of possible sizes and densities of the lunar core. In order to test the plausibility of these lunar interior models, we calculated the bulk Moon density and BSM moment of inertia for each model and compared the results with the observed physical properties of the Moon (see table 1).

## 4.1. Mixing models

As detailed in section 3, we simplified the complex compositional layering of the LMO cumulate by discerning 5 reservoirs: a dunitic lower mantle (LM), a pyroxenitic upper mantle (UM), dense Ti-rich cumulates (IBC), late forming light cumulates (KREEP) and an anorthositic crust. The relative masses and exact compositions of these primary reservoirs depend on the assumed compositional model (i.e. the FeO/MgO ratio in the assumed bulk LMO composition). All primary reservoirs are assumed to be chemically homogeneous with a bulk composition corresponding to the average composition of all individual layers from the fractional crystallization model that were combined to form the reservoir. To simulate mixing and chemical equilibration of the mantle reservoirs, we merged the LM and UM reservoirs into a single mantle reservoir ("mixed UM+LM" in Fig. 3) or mixed LM, UM and IBC ("homogeneous mantle" in Fig. 3). We assumed that these mixed layers obtained complete chemical equilibration for the calculation of their physical properties, which is further detailed below. KREEP and crustal material are not considered to have taken part in mantle convection and mixing due to their low densities. Furthermore KREEP has a very small volume compared to the other reservoirs, so that the effect of any displacement of KREEP on the mass distribution in the BSM is negligible.

## 4.2. Stratigraphic models

The primary or mixed compositional reservoirs described above were arranged in different stratigraphic configurations (Fig. 3) to simulate different mantle overturn scenarios. Models 1-4 assume a moderately mixed mantle with separate LM and UM reservoirs. Model 1 represents the original layer configuration without any stratigraphic changes by convection during or after LMO solidification. Model 2 assumes that all IBC material was transported to the core mantle boundary (CMB) without disturbing the stratigraphy of the other layers. This might not be a realistic stratigraphic configuration for the lunar mantle, but it has been included to study the isolated effect of IBC sinking by comparing models 1 and 2. Models 3 and 4 both assume that the large mantle reservoirs have completely overturned (without mixing), but differ in the position of the IBC layer, which is stuck below the KREEP layer in model 3 and has been transported to the core mantle boundary in model 4. Models 5 and 6 assume a strongly mixed mantle where the UM and LM reservoirs have been merged into one homogeneous layer. Model 5 assumes that IBC did not sink, while model 6 assumes all IBC sunk towards the CMB. Model 7 finally assumes that IBC, UM and LM have been completely homogenized by mixing. In all models KREEP is assumed to remain at its original position beneath the lunar crust. Even if some KREEP material was entrained by deeper mantle layers, the KREEP reservoir represents less than 0.5 wt% of the bulk silicate Moon, so that its position does not significantly affect the bulk Moon properties.

### 4.3. Mass distribution in today's lunar mantle

In order to test the consistency of different lunar interior models with geophysical properties, we calculated the BSM density and moment of inertia for each model, considering the masses and depth dependent densities of the respective layers. Depth dependent material densities were calculated beforehand with Perple_X (Connolly, 2005) for each material appearing in the stratigraphic models, considering their respective bulk compositions and the change of local pressures and temperatures with depth. This calculation requires assumptions regarding the thermal structure of the lunar interior. Additional assumptions about the mass and volume of the lunar core have to be made to relate the calculated BSM density to the observed bulk Moon density. Therefore, we assumed a range of different core models and selenotherms, which are discussed in the following sections. All calculations of material densities were made for different bulk silicate Moon FeO contents of 8 – 13 wt%, resulting in different relative masses and compositions of the different layer materials.

#### *4.3.1. Thermal models of the lunar interior*

Several studies have proposed hot selenotherms with core-mantle boundary temperatures around 1500-1600 K (e.g. Gagnepain-Beyneix et al., 2006; Khan et al. 2014, Laneuville et al. 2013). Such high temperatures are consistent with the liquid state of the outer core (Weber et al., 2011, Garcia et al., 2011). Furthermore, the attenuation of seismic waves in the deep lunar interior and the observed tidal dissipation indicate the presence of a low viscosity zone at the CMB that is consistent with partial melting of the lowermost mantle (e.g. Weber et al., 2011; Harada et al., 2014). Partial melting of the lowermost mantle can only be expected for temperatures exceeding ~1600 K (Mallik et al., 2019).

However, these high temperatures are difficult to reconcile with the existence of deep moon quakes which suggest a brittle state for the lower mantle (Kawamura et al., 2017). Kawamura et al. (2017) re-evaluated the brittle-ductile transition temperature for large tidally induced strain rates and found that a brittle lower mantle is consistent with cool selenotherms with core-mantle boundary (CMB) temperatures of ~1273 ±100 K, similar to those proposed by Kuskov and Kronrod (1998). Considering a pyroxene bearing lower mantle instead of a pure olivine lower mantle could further increase the brittle-ductile transition temperature by up to ~95 K. Thus, according to this model ~1470 K is the maximum lower mantle temperature that is consistent with the occurrence of deep Moon quakes at up to 1200 km depth.

Since both constraints of a partially molten lowermost mantle and brittle behavior of the lower mantle are incompatible, we assume two different selenotherms with different lower mantle temperatures (Fig. 4). For the colder selenotherm we assume the thermal profile proposed by Kuskov and Kronrod (1998) (their model 1). As representative for a possible hot selenotherm consistent with partial melting at the CMB. we assume the thermal profile proposed by Laneuville et al. (2013) for the lunar Farside.

*4.3.2. Core models*

The size and density of the lunar core are critical parameters in our model, which influence the bulk Moon density and the volume of the bulk silicate Moon. However, the current physical properties of the lunar core are still poorly constrained. Estimates of the lunar core radius based on seismic measurements lie in a range of about 310 – 420 km (e.g. Garcia et al., 2011; Weber et al., 2011), while magnetic field data analyses indicate that the lunar core radius does not exceed 400 km (Shimizu et al., 2013). Recent lunar laser ranging data restrict the lunar core size more precisely to a radius of 381 ± 12 km (Viswanathan et al., 2019).

To date there is no clear consensus about whether the lunar core is completely molten (e.g. Garcia et al., 2011) or possesses a solid interior (e.g. Weber et al., 2011), which complicates estimates of the bulk core density. Bulk core density estimates range from ~4200 – 7400 kg/m³ (Garcia et al., 2011; Antonangeli et al., 2015), depending on the existence and size of an inner core as well as the core temperature and the fractions of siderophile elements like S or Ni, that influence the density and melting temperature of the lunar core.

To make sure that we cover the whole range of plausible core properties, we assume a range of different core models with radii ranging from 330 – 430 km and densities of 4000 – 7500 kg/m³. The properties of these core models are illustrated in Fig. 5. They include both literature estimates of core properties (Garcia et al., 2011; Weber et al., 2011; Antonangeli et al., 2015) and an additional set of core models that we calculated assuming a core size of 381 ± 12 km (Viswanathan et al., 2019), a bulk core S content of 8 wt% and core densities consistent with the assumed CMB temperatures and the composition dependent densities for partially crystallized Fe-S alloys reported by Antonangeli et al. (2015). Details of these calculations are given in the supplementary material.

*4.3.3 Calculation of bulk Moon density and BSM moment of inertia*

To determine the bulk Moon density and BSM moment of inertia for a given lunar interior model, we divided each compositional reservoir into 10-20 sublayers of equal volume. The initial volumes of the layers correspond to the volumes at the time of their formation as calculated by the crystallization model and were later adjusted to be consistent with the selected present day selenotherms and stratigraphies.

For each sublayer, we calculated the average density based on the depth-density function of the respective material and the depth of the upper boundary of the sublayer. The updated density was used to re-calculate the volumes of the individual layers while conserving their mass proportions. In a next step the layer volumes were normalized as to fit the bulk Moon volume in combination with the chosen core size. The new layer depths were then used to update the layer densities in the next iteration step and the procedure was repeated until the layer densities were stable and the resulting bulk Moon radius fit the real value within an error of <1km.

The chosen number of sublayers was sufficient to ensure that the density changes between neighboring sublayers were always smaller than 10 kg/m³. The error introduced by assuming a pressure and temperature (and corresponding density) based on the upper boundary depth rather than the lower boundary depth turned out to be negligible for both the BSM moment of inertia and the bulk Moon density for the used number of sublayers.

## 5. Physical properties of lunar interior models

Our lunar interior models consider variations in several parameters that affect the calculated bulk Moon density and BSM moment of inertia. In the following we discuss the systematics of these effects and how they can be used to derive systematic relations between different properties of the lunar interior.

### 5.1. Effect of core models on bulk Moon physical properties

Fig. 5 illustrates the effect of varying core sizes and densities on the calculated bulk Moon density and BSM moment of inertia. All depicted models assume a BSM composition with 9 wt% FeO, a cold selenotherm and an interior structure according to the stratigraphic model 1.

As illustrated in Fig. 5a,c, the bulk Moon density increases both with increasing core radius and increasing core density. However, changing either core radius or core density produces two different trends (blue and yellow lines in Fig. 5a-f). This is because changing the core density while keeping the core volume constant affects only the mass of the core, while the mass and volume of the BSM remain constant. Changing the core volume while keeping the core density constant on the other hand affects not only the core mass, but also the volume of the BSM. This is because the bulk Moon volume is a fixed quantity, so that any increase in core volume must be compensated by a decrease in BSM volume. Due to these different effects of changing core radius and core density, there is no simple relation between bulk Moon density and assumed core mass, though generally bulk Moon density increases with core mass (Fig. 5e).

Since the BSM moment of inertia does not include the moment of inertia of the core, the mass and density of the core do not affect the BSM moment of inertia. The apparent dependence of the BSM moment of inertia on the core mass shown in Fig. 5f is only an indirect effect of higher core masses being typically associated with larger core volumes. Larger core volumes are associated with smaller BSM volumes and masses – and hence a smaller BSM moments of inertia.

Overall, the effect of the core properties on the BSM moment of inertia factor is small (± 9.5·10$^{-5}$) compared to the effect of FeO contents or assumed mantle stratigraphies (see Fig. 6), but still substantially larger than the uncertainty of the measured BSM moment of inertia factor (± 1.2·10$^{-5}$, Matsumoto et al., 2015).

**5.2. Effect of stratigraphic models on bulk Moon physical properties**

Fig. 6 shows the bulk Moon density and BSM moment of inertia for the different lunar interior models illustrated in Fig. 3., assuming a hot selenotherm (lunar Farside model, Laneuville et al., 2013) and FeO contents of 9 wt% (a) and 13 wt% (b). The colors and color depths symbolize different stratigraphic and mixing models, and the dots along each line correspond to different core properties assumed in the respective models.

Generally, the different stratigraphic models have very similar ranges of bulk Moon density values, since the variation of bulk Moon density is dominated by the assumed properties of the core and the chosen BSM FeO content. Stratigraphic models without overturn of the large mantle reservoirs LM and UM (models 1, 2, shown in purple) have slightly lower densities than the models with overturned LM and UM reservoirs (models 3, 4, shown in blue). The higher densities in models 3 and 4 can mainly be attributed to the formation of high-density phases (garnet) in the Al-rich UM material as it is transported to greater depths during mantle overturn and experiences higher pressures at its final position. The LM and IBC materials on the other hand do not experience any significant phase changes during ascent/sinking and the density variations due to pressure and temperature changes are subtle, so that the convective transport of these materials does not noticeably affect bulk moon density.

However, the distribution of mass in the lunar interior differs strongly for the different stratigraphic models, leading to distinct differences in the BSM moment of inertia. The highest values in the BSM MoI are reached for a lunar interior that did not experience any changes by solid state convection but preserves the gravitationally unstable density structure produced by bottom up LMO crystallization (model 1, light purple). Lower values of BSM MoI are obtained for models where denser material has been transported to greater depths by overturn of the original stratigraphy (models 3 and 4, blue) or by mixing of layers (models 5 and 6, green). Model 4 (dark blue), which represents a completely overturned mantle but without mixing of the layers, has the lowest value of BSM MoI since this model features the strongest increase in density towards the core. Due to the high density of IBC material, the position of IBC has a large effect on the BSM moment of inertia, which becomes apparent when comparing pairs of stratigraphic models that differ only in the position of IBC (i.e. models with the same color but different color depth in Fig. 6). The magnitude of this effect depends on the assumed BSM FeO content, as will be discussed in more detail below.

**5.3. Effect of FeO content on bulk Moon physical properties**

The BSM FeO content systematically affects the BSM moment of inertia and the bulk Moon density. As illustrated in Fig. 7, both quantities increase linearly with the FeO content for a given stratigraphic model. This systematic trend allows us to interpolate between individual experimental datasets and to define an empirical function for the change of density and moment of inertia with FeO content for each stratigraphic model.

As shown in Fig. 2, increasing BSM FeO contents lead to systematically increasing volumes of IBC material. Due to the high density of IBC, its position has a strong effect on the BSM moment of inertia (Fig. 6). The magnitude of this effect depends on the volume of IBC and hence on the BSM FeO content. When comparing pairs of stratigraphic models in Fig. 6a and 6b that only differ in the position of IBC (models 1 and 2, models 3 and 4 or models 5 and 6, respectively) it becomes apparent that the effect of IBC sinking on the BSM moment of inertia changes systematically with the BSM FeO content. This dependency is the same for all types of models, i.e. the difference in BSM moment of inertia between models 1 and 2 is the same as between models 3 and 4 or between models 5 and 6. Hence the magnitude of the effect of IBC sinking on the BSM moment of inertia as a function of BSM FeO content can be described by an empirical function that interpolates between the model data.

In the following we apply these empirical functions to test the consistency of the lunar interior models with the bulk Moon density and BSM moment of inertia and discuss to what extent our results can be applied to constrain the BSM FeO content.

**5.4. Quantification of the interrelations between model parameters**

In the previous section we have identified four systematic effects of the BSM FeO content:

(1) A systematic change of bulk Moon density with the BSM FeO content (Fig. 7).

(2) A systematic change of BSM moment of inertia with the BSM FeO content (Fig. 7).

(3) A systematic change of IBC thickness with the BSM FeO content (Fig. 2).

(4) A systematic change of the total variance of the BSM moment of inertia induced by IBC sinking with the BSM FeO content (Fig. 6) resulting from (3).

These relations can be quantified by defining simple empirical functions that interpolate between the model data. However, these functions need to be defined specifically for every type of stratigraphic model and selenotherm, since the BSM moment of inertia and bulk Moon density also depend on the mantle stratigraphy and temperature.

The first two relations can be used to determine which BSM FeO contents need to be assumed to obtain realistic BSM moment of inertia and bulk Moon density values, given a specific stratigraphic model and selenotherm. However, the 7 stratigraphic models we considered are end member cases and do not cover the complete parameter space of possible bulk Moon densities and BSM moments of inertia that could be realized by partial (rather than either complete or absent) mixing and overturn of primary compositional reservoirs. Therefore, it is useful to introduce another free parameter that describes the range of changes in BSM moment of inertia and/or bulk Moon density produced by changes in the stratigraphy. As described by the third and fourth relation listed above, the thickness and position of IBC are stratigraphic parameters that are strongly connected to both the BSM FeO content and moment of inertia and are hence useful for both the characterization of intermediate stratigraphies and our objective of associating the BSM FeO content with physical properties of the BSM. The positions of the LM and UM reservoirs on the other hand affect both the bulk Moon density and BSM moment of inertia, but this effect is almost independent of the BSM FeO content. Therefore, we chose the distribution of IBC as parameter to interpolate between the BSM moment of inertia and bulk Moon density values of different stratigraphic models.

The increase of IBC thickness with BSM FeO content follows a linear systematic that is easy to interpolate (see Fig. 2). The connection between the radial position of IBC and the BSM moment of inertia is less straightforward. In the end member stratigraphic models 1-6 all IBC material is bundled in a single layer that is positioned either at its original position of formation or at the CMB. Intermediate states of IBC distribution can generally occur if a) only a part of IBC layer sinks to the CMB while the rest remains at the original position or b) some IBC sinks only a part of the distance towards the CMB, because it gets mixed into the mantle and stays there. Thus, for a given fraction of IBC that sinks into the mantle there are several possibilities how the sunken IBC can be finally distributed – and these different distributions would result in different BSM moments of inertia. Hence the fraction of sunken IBC is not a feasible parameter to represent the effects that intermediate states of IBC distribution have on the BSM moment of inertia. For the interpolation of the moment of inertia and density space between the end member stratigraphic models we therefore used the "normalized change in the BSM moment of inertia that is caused by IBC sinking" as additional free parameter instead of using the more intuitively accessible "fraction of sunken IBC".

The absolute change in the BSM moment of inertia caused by IBC sinking corresponds to the vertical distance of same-colored model lines in Fig. 6. Since this absolute change in the BSM moment of inertia depends on the (BSM composition dependent) IBC thickness (see Fig. 6 a, b), we use normalized values where a value of 0 represents a model with IBC at the original position (no change in the BSM moment of inertia by IBC sinking) and a value of 1 corresponds to a stratigraphy where all IBC is at the CMB (maximum change in the BSM moment of inertia by IBC sinking).

The "normalized change in the BSM moment of inertia caused by IBC sinking" can be related to the fraction of sunken IBC by introducing a factor that weighs each increment of sunken IBC based on its radial position and related effect on the BSM moment of inertia. In the following we will hence refer to the "normalized change in the BSM moment of inertia caused by IBC sinking" as "weighted fraction of sunken IBC" in order to use a shorter and more intuitively accessible term in favor of a better readability.

The interpolation between model pairs that differ only in their IBC position produces additional sets of stratigraphic models that can be grouped based on the positions of the LM and UM layers in the mantle. This way we obtain four types of mantle stratigraphies based on our 7 original stratigraphic models:

1) LM and UM in original position (derived from interpolation between models 1 and 2, purple in Fig. 6 and 8),

2) LM and UM overturned (derived from interpolation between models 3 and 4, blue in Fig. 6 and 8),

3) LM and UM homogeneously mixed (derived from interpolation between models 5 and 6, green in Fig. 6 and 8) and

4) the original model 7 with a homogeneous mixture of LM, UM and IBC material (yellow in Fig. 8).

Based on the functions describing the relations between BSM moment of inertia, bulk Moon density, BSM FeO content and the weighted fraction of sunken IBC, we developed an algorithm that for each of the 4 types of stratigraphic models

a) enters a loop that goes through a user-defined range of BSM FeO contents

b) calculates the BSM moment of inertia and bulk Moon density values for the two end members of the stratigraphic model (with IBC at the position of formation or at the CMB)

c) interpolates between the two end member models to determine which weighted fraction of sunken IBC needs to be assumed to reproduce the actual bulk Moon density and BSM moment of inertia values reported by Matsumoto et al. (2015)

d) checks if the determined value for the weighted fraction of sunken IBC is physically plausible (i.e. a value between 0 and 1)

e) records the BSM FeO contents and weighted fractions of sunken IBC for all lunar interior models that fit the actual bulk Moon density and BSM moment of inertia values, considering the uncertainty in the observed values of bulk Moon density and BSM moment of inertia as reported by Matsumoto et al. (2015) (see table 1).

We applied this algorithm to all 4 types of stratigraphic models, assuming both the hot and the cold selenotherm shown in Fig. 4. The results of these calculations are summarized in Fig. 8 and are described in more detail in the following section.

## 5.5. Consistency of lunar interior models with bulk Moon physical properties

The combinations of BSM FeO content and weighted fractions of sunken IBC that are consistent with the actual bulk Moon density and BSM moment of inertia are shown in Figure 8. The colored areas in the figure represent the four types of mantle stratigraphies described in the previous section, which are end member cases that represent different positions and degrees of mixing of the LM and UM mantle reservoirs. Since intermediate cases between those stratigraphies exist, the space between the colored areas is also a valid parameter space for the lunar interior.

Since the bulk Moon density is a function of temperature, the range of possible FeO contents for a given stratigraphic model depends also on the assumed selenotherm. Higher temperatures are generally associated with lower material densities, so hotter selenotherms require higher BSM FeO contents (see Fig. 8 a and b).

The range of possible FeO contents for each type of interior model is limited along the x-axis by the range of moment of inertia values produced by IBC sinking and along the y-axis by the consistency of the assumed core models with the bulk Moon density.

For a given lunar interior model, the weighted fraction of sunken IBC is positively correlated with the BSM FeO content (Fig. 8). This is because the BSM FeO content is directly related to the initial IBC layer thickness: The higher the FeO content, the thicker the initial IBC layer and the higher the fraction of IBC that needs to sink to fit the BSM moment of inertia. At high FeO contents the BSM moment of inertia depends more strongly on the distribution of IBC (since the total IBC mass is larger). As a consequence, the colored zones in Fig. 8 are slightly curved upwards and the slope of the curves is steeper in hotter models that have generally higher BSM FeO contents.

Although all distributions of IBC (i.e. weighted fractions of sunken IBC) were considered to be possible in the model, some of the colored zones in Fig. 8 do not cover the whole range of the diagram. These truncations are a consequence of limitations in the range of core properties we assumed (see table 1 for the used core model properties). High BSM FeO contents are associated with high BSM densities, that need to be balanced with a low core density and/or a small core size to fit the bulk Moon density. As illustrated in Fig. 5, the bulk Moon density is affected by both core density and core volume and changes in both quantities produce different independent trends, so that there is no simple correlation of FeO content or fraction of sunken IBC with the core mass (see discussion in section 5.1).

In Fig. 8 black areas connected by gray lines represent the parameter space that was calculated for specific core models from the literature (Garcia et al., 2011; Weber et al., 2011; Antonangeli et al., 2015) in combination with different types of mantle stratigraphies. It should be noted that these core models were constructed assuming high core temperatures (~1650 K or higher). Therefore a combination of these core models with cold selenotherms is not realistic and is only shown in Fig. 8 for comparison with the hot selenotherm case.

The gray lines signify the parameter space that would be covered from the respective core models if they were combined with intermediate mantle stratigraphies (i.e. stratigraphies with partial mixing and/or overturn of mantle layers) that were not explicitly modeled in this study. The properties of the shown core models and the associated ranges of BSM FeO contents and weighted fractions of sunken IBC corresponding to the gray lines in Fig. 8 are listed in table 2.

The gray lines have a small positive slope, because higher FeO contents generally lead to increasing IBC masses and require a larger fraction of sunken IBC to fit the BSM moment of inertia. However, the range of possible FeO contents for a given core model is very limited, because the core model strongly affects the bulk Moon density and thus limits the range of possible BSM densities (which in turn mainly depend on FeO content). This means that tighter constraints on the properties of the lunar core would a) allow tighter constraints on the BSM FeO content and b) define a tight relation between the weighted fraction of sunken IBC and the structure and degree of mixing of the rest of the mantle, which could be especially useful in combination with further independent constraints on the interior structure.

The red area marks the parameter space defined by the core models c1-6, which have been constructed from LLR based core size estimates by Viswanathan et al. (2019) and densities consistent with a bulk core S content of 8 wt% (Antonangeli et al., 2015) and the CMB temperatures of the respective selenotherms. This parameter space and the associated structure models can be considered as most realistic for the lunar interior.

## 6. Discussion

**6.1. Constraints on bulk silicate Moon FeO contents**

Our models show that the BSM FeO content can be constrained by considering its effects on the densities and relative proportions of mantle reservoirs that form by LMO solidification and the possible effects of convection on the mantle stratigraphy and the corresponding BSM moment of inertia. For a cold lunar interior (selenotherm after Kuskov and Kronrod, 1998), the calculated BSM FeO contents range between 8.0 and 11.2 wt%, while for a hotter lunar mantle (Lunar Farside model by Laneuville et al., 2013) the BSM FeO contents are between 9.4 and 12.7 wt%. These estimates of the BSM FeO content can be narrowed down further if we consider more information about the mixing behavior of the lunar mantle, the thickness of the IBC layer at the core-mantle boundary and the size and density of the lunar core.

*6.1.1. Constraints from core size and density*

The general trends shown in Fig. 8 are constructed for a large range of core sizes and densities, which can be narrowed down to more realistic values considering seismic and selenodetic data as well as petrological arguments regarding core composition. Recent LLR data suggest that the lunar core has a radius of 381±12 km (Viswanathan et al., 2019), which is consistent with the range of core radii suggested in earlier studies based on seismic data (380 ±40 km Garcia et al., 2011; 330 ±20 km Weber et al., 2011).

The density of the core is much more difficult to constrain, because neither the temperature nor the composition of the lunar core are precisely known and the size of the solid inner core is still poorly constrained. Seismic studies indicate that the lunar core is at least partially molten and suggest a large range of bulk lunar core densities of ~4200 – 7400 kg/m³ (Garcia et al., 2011; Antonangeli et al., 2015). Such comparatively low densities require the presence of light elements like S or C in the core. Antonangeli et al. (2015) estimated an outer core thickness of 80-85 km and concluded from molten metal alloy density and liquidus considerations that the core should contain 3-6 wt% light elements, while other studies could not resolve an inner core boundary using the available seismic data (Garcia et al., 2011). Based on the siderophile element contents of the lunar mantle, recent studies suggest core S contents ranging from < 0.16 wt% to 6wt% (Steenstra et al., 2017; Rai and van Westrenen, 2014) and C contents of up to 4.8 wt% (Steenstra et al., 2017), while core dynamo models suggests light elements contents of 6-8 wt% in the lunar core (Laneuville et al., 2014). The range of outer core densities associated with these light element contents can be estimated using experimental data on liquid Fe-S densities (Antonangeli et al., 2015), since the densities of Fe–C and Fe–S melts are very similar (Shimoyama et al. 2013).

Estimating the bulk core density further requires constraints on the degree of melting. The melting temperatures of the core are affected by the concentrations of S, C and Ni in the core, so that we can use respective phase diagrams (Liu and Li, 2020) to estimate which degree of core solidification is realistic for a given CMB temperature. Estimates for the core Ni content range from 8–9 wt% (Righter and Drake, 1996; Righter et al., 2017) to 35–55 wt% (O'Neill,1991). A fully molten core would require 22-23 wt% S at 1236 K (our cold selenotherm) and about 8-11 wt% S at 1683 K (our hot selenotherm). This is a much higher S content than it is expected from petrological arguments. For a S content of 8 wt% a fully molten core requires a temperature of at least 1675-1750 K depending on the Ni content of the core (Liu and Li, 2020). Therefore it seems likely that the core is partially crystallized.

The core models c1-6 (Fig. 8) consider our preferred estimates for the core size (381±12 km, Viswanathan et al., 2019) and densities consistent with the CMB temperatures of the assumed selenotherms and a light element content of 8 wt%. By assuming such a high concentration of light elements we obtain lower limits for plausible core densities and upper limits for the associated BSM FeO contents. To include an absolute lower limit for the FeO content, we also assumed a completely solid iron core (marked as csolid in Fig. 8). The degree of solidification varies for these cores from 0-29% for the hot selenotherm and 64-65% for the cold selenotherm. If the thickness of the outer core layer is indeed 80-85km as determined by Antonangeli et al. (2015), then the temperature of the lunar interior needs to be at least 1525 K assuming 8wt% S in the core, or higher if the S content is lower.

In the cold selenotherm case, the core models c1-6 constrain the BSM FeO content to 8.1 – 9.2 wt %, while the minimum possible FeO content considering lower light element concentrations is 8.0 wt%. In the hot selenotherm case the range of plausible BSM FeO contents is 9.9 – 11.8 wt%, with a lowermost limit of 9.5 wt% FeO if the core was completely solid and free of any light elements. A completely molten core can be excluded for the cold selenotherm case, while the hot selenotherm case allows a completely molten core, but only at both high light element concentrations of at least 8 wt% and high bulk core Ni contents of Ni/(Ni+Fe) > 0.55.

*6.1.2. Constraints on the mantle stratigraphy from seismic data*

An important source of information on the thermal and compositional structure of the lunar interior are density and seismic velocity models based on the arrival times of seismic signals. Lognonne et al. (2003) and Gagnepain-Beyneix et al. (2006) proposed several possible pyroxenitic upper mantle compositions and a Mg-rich lower mantle with magnesium numbers increasing with depth. Their first pyroxenitic upper mantle model is based on the source composition of mare basalts (Ringwood and Essene, 1970) and is consistent with a mixture of our pyroxenite reservoir with about 6% trapped plagioclase (which is consistent with a ~31 km thick plagioclase crust in our model), derived from an FeO-rich (~12.5 wt%) magma ocean. The second pyroxenitic upper mantle model proposed by Gagnepain-Beyneix et al. (2006) (after Kuskov, 1995) is consistent with a mixture of our pyroxenitic reservoir with all available IBC material, also considering an FeO-rich (~12.5 wt%) magma ocean. Both of these upper mantle compositional models suggest that the overturn of the largest mantle reservoirs was incomplete, since the structure of pyroxenite overlying a dunitic lower mantle corresponds to the original cumulate layering produced during magma ocean solidification. Regarding the sinking of IBC material the results are inconclusive with one model requiring all available IBC material in the upper mantle while in the other model IBC material is absent in the upper mantle.

According to our results, a lunar mantle consisting of a dunitic lower mantle and a pyroxenitic upper mantle as proposed by Gagnepain-Beyneix et al. (2006) should be associated with low BSM FeO contents (<11.5 wt% for the hot and < 10.1 wt% for the cold selenotherm) to fit the bulk lunar density and BSM MoI (purple bands in Fig. 8). However, the BSM FeO contents that are required to form the reservoirs proposed by Gagnepain-Beyneix et al. (2006) by magma ocean fractional crystallization are distinctly higher (12.5 wt% FeO) than would be consistent with our model. A related discrepancy between our model and that of Gagnepain-Beyneix et al. (2006) is the composition of the lower mantle. We can generally construct their proposed Mg-rich lower mantle composition by mixing our olivine-dominated reservoir with 14% IBC and 9% crustal material. However, considering the relative masses of these reservoirs, such a mixture would require higher amounts of IBC and crustal material than are available based on the magma ocean crystallization model. This indicates that higher amounts of Ca, Al and Fe need to be assumed for the BSM to fit the proposed mantle reservoir composition. Such changes would increase the BSM density. Both Lognonne et al. (2003) and Gagnepain-Beyneix et al. (2006) note that the compositional models they propose produce bulk mantle densities that exceed the bulk Moon density and are hence not realistic. However, regardless of the exact composition, their results indicate that the seismic velocities in the lunar mantle tend to increase with depth, which is generally consistent with a more pyroxenitic upper mantle and a dunitic lower mantle. This stratigraphic configuration suggests that the overturn of the dunitic and pyroxenitic reservoirs in the lunar interior was limited.

*6.1.3. Constraints on the mantle stratigraphy from dynamical models*

As discussed above, seismic models suggest that overturn of the dunitic and pyroxenitic reservoirs in the lunar interior was limited (Lognonne et al., 2003; Gagnepain-Beyneix et al., 2006). Dynamical models, however, indicate that cumulate convection – and hence mixing and overturn of mantle reservoirs – might occur already before the magma ocean has fully crystallized (e.g. Maurice et al., 2017). This discrepancy can possibly be explained if partial melting is considered. Convection within the cumulates can lead to compositional mixing, but also to partial melting of the mantle material (Maurice et al. 2020). Partial melting leads to further fractionation into a cumulate that is depleted and a melt that is enriched in incompatible elements. This fractionation process counteracts the simultaneous convective mixing of different compositional reservoirs. Hence efficient cumulate convection might not necessarily result in a well-mixed mantle, depending on whether mixing or fractional melting dominate the compositional evolution. A final evaluation and quantification of the compositional effects of cumulate melting during convection in the lunar interior is beyond the scope of this paper but would certainly be worth pursuing in future studies.

### 6.1.4. Constraints from dynamical models of IBC sinking

In our models we made the simplifying assumption that the IBC can be located either at the base of the crust and/or at the core mantle boundary – apart from the strong mixing model where IBC is homogeneously mixed into the mantle. However, dynamical models of IBC sinking indicate that IBC material sinks in the form of small diapirs, which results in some mixing of IBC material into the mantle (e.g. Hess and Parmentier, 1995; Yu et al., 2019). Recent models by Yu et al. (2019) suggest that if nearly all IBC material sank into the mantle, only about 50% of the IBC material would accumulate at the core mantle boundary, while the remaining IBC material would get homogeneously mixed into the mantle. This behavior has implication for the range of BSM moment of inertia values that can be produced by IBC sinking.

If IBC sinking is always accompanied by some mixing of IBC material into the mantle, the end member stratigraphies with all IBC at the CMB can never be realized. This dynamical constraint limits the range of plausible BSM FeO contents, because stratigraphies with high weighted fractions of sunken IBC are associated with high FeO contents (Fig. 8). Though current dynamical models do not explicitly address the radial distribution of IBC as a function of the fraction of sunken IBC, such a relation could in principle be established using current dynamical modeling tools.

### 6.1.5. Constraints on the amount of IBC at the CMB from selenodetic data

Matsumoto et al. (2015) combined current selenodetic and seismic data to constrain the density structure of the lunar mantle without explicitly fitting material compositions and temperatures.

Their results indicate the presence of a 170 km thick zone at the core mantle boundary that is characterized by low seismic velocities and high densities of 3450 – 3650 kg/m³. These densities are too high for typical mantle material (UM and LM both have densities < 3420 kg/m³), but could be consistent with the presence of IBC material. Pure IBC material has densities of ~4000 kg/m³ at the pressures and temperatures close to the CMB, so that the low velocity zone (LVZ) density can only be fitted by a mixture of mantle and IBC material.

Since the amount of IBC material mixed in the low velocity zone can be calculated if the densities of mantle and IBC material are known, we can use the constraints on the LVZ thickness and density to calculate the corresponding fraction of IBC that reached the CMB for a given lunar interior model. As discussed above, dynamical models indicate that any sinking IBC material gets partially mixed into the mantle and that only about half of the IBC material that sank into the mantle might finally reach the CMB. If we combine this assumption for the distribution of sunken IBC with the constraints for the LVZ thickness, we can calculate the range of plausible weighted fractions of sunken IBC for different lunar interior models. These ranges are marked in Fig. 8 by the areas bordered with red dotted lines.

### 6.2. Estimates of BSM FeO content and weighted fractions of sunken IBC

Considering the constraints from the LVZ thickness and density in combination with the constraints on mantle stratigraphy and core properties discussed above results in our preferred lunar interior model that limits the BSM FeO content to 8.3 – 9.0 wt% for the cold selenotherm and 10.0 – 10.9 wt% for the hot selenotherm (preferred model encircled by black lines in Fig. 8). The associated weighted fractions of IBC range from 19 – 61% (cold selenotherm) to 22 – 58% (hot selenotherm). Since these estimates include the assumption of comparatively high core S contents, the lower limits imposed by the core properties (red area in Fig. 8) cannot be considered as hard constraints. Therefore the FeO contents might be lower in the hot selenotherm case where the lower limit of the preferred model is defined by the core properties.

Assuming a core S content of 8 wt% and an outer core thickness of 80 – 85 km (which requires core temperatures of about 1525 – 1560 K) restricts the BSM FeO contents to 9.4 – 10.4 wt% and the weighted fractions of sunken IBC to 21 – 59%. The same liquid fraction of the core could also be reached by assuming lower core S contents and higher temperatures. The CMB temperature of the hot selenotherm (1683 K) requires S contents of 4 – 6 wt% to reach outer core thicknesses of 80 – 85 km. This corresponds to the lower limit of light elements required in the core given by Laneuville et al. (2014) based on core dynamo models.

Hence, considering all available constraints, our models suggest that core temperatures with 1525 – 1683 K, BSM FeO contents of 9.4 – 10.9 wt% (with a lowermost limit of 9.0 wt%) and weighted fractions of overturned IBC of 21 – 59 % are most realistic for the lunar interior. Warren (2005) estimated similar BSM FeO contents of 8.3 – 10.4 wt% FeO (preferred model 9.1 wt% FeO) based on the compositions of lunar rocks and considerations regarding cosmochemically plausible main and trace element ratios.

### 6.3. Limitations and applicability of the model

*6.3.1. BSM Composition*

In this study we have only considered a limited range of possible BSM compositions by considering only the effect of changing FeO/MgO ratios. Though all magma ocean compositions proposed to date result in similar crystallization sequences, variations in the abundances of other oxides would affect the density and mineralogy of different cumulate layers. Higher Al and Ca contents can e.g. be expected to lead to larger amounts of garnet in the lunar mantle and hence to increase mantle densities. Such higher mantle densities would shift the range of possible BSM FeO contents towards lower values. In general, the modeling approach can also be applied to Al-Ca richer BSM compositions, but would require the validation of the LMO crystallization model by experiments in this compositional range (e.g. Charlier et al. 2018).

*6.3.2. Lunar interior temperature*

The 1D models used in this study consider only radial temperature variations and hence do not include any lateral temperature variations like the Nearside-Farside dichotomy. The hot selenotherm used in this work corresponds to the temperature distribution for the Lunar Farside proposed by Laneuville et al. (2013). The Lunar Nearside selenotherm proposed by Laneuville et al. (2013) includes significantly hotter temperatures at shallow depths. In our bulk Moon models such a selenotherm would be associated with significantly higher BSM FeO contents (14.4 – 17.2 wt%), but we do not consider these values as realistic since the temperature anomaly on the Lunar Nearside is not representative for the bulk Moon. However, considering elevated temperatures on the Lunar Nearside would slightly increase the laterally averaged temperatures of our selenotherms and shift the resulting BSM FeO contents to higher values. Assuming that the temperature anomaly at the Lunar Nearside extends over approximately 1/6 of the lunar surface, a consideration of the temperature anomaly would increase our estimates of the BSM FeO content by ~1wt%.

*6.3.3. Magma ocean depth*

In our models we assume a deep magma ocean that comprises the whole bulk silicate Moon. If the BSM was only partially molten, it would comprise both magma ocean cumulates and a primitive mantle. The presence of such a primitive mantle would introduce an additional compositional reservoir, which could potentially affect the range of possible densities and moments of inertia for a given BSM composition.

The total range of possible BSM moments of inertia depends on how strong the density differences between different mantle layers are. Since magma ocean solidification leads to the differentiation of the mantle into layers of different composition and density, the possible range of BSM moments of inertia is the higher, the deeper the magma ocean is.

The range of densities depends strongly on the types of phases that can be formed in the different mantle reservoirs. Differentiation by magma ocean solidification leads to high concentrations of Ca and Al in the late forming cumulate layers and a complementary depletion in the early cumulate layers. The concentration of Ca and Al plays an important role for the density of lunar mantle layers, because high concentrations of Ca and Al facilitate the formation of dense garnet at elevated pressures. Therefore Ca-Al-rich layers produced by LMO solidification can experience a significant increase in density when they are transported to larger depth, while Ca-Al-poorer layers (the primitive mantle or depleted LMO cumulates) experience a smaller change in density when they are transported to a different depth. Hence, in a scenario with a deep magma ocean, there is a larger range of possible BSM densities – given that the layers can be arranged in different configurations as we assume it in our models.

Following these considerations, we can conclude that a shallower magma ocean would lead to smaller possible ranges of BSM density and moment of inertia for a given BSM composition and consequently stronger constraints on the BSM FeO content. Our choice of assuming a deep magma ocean hence allows us to keep our estimates of the BSM FeO content as reliable as possible by including all possible scenarios of LMO extent in the range of estimated values.

*6.3.4. Degree of crystal fractionation in the magma ocean*

The degree of crystal fractionation determines the efficiency of the chemical exchange between the cumulate layers and the coexisting liquid. A high degree of fractionation is associated with a large diversity in cumulate layer compositions, while a lower degree of fractionation leads to smaller compositional differences among the cumulate layers. As discussed above, a larger diversity in mantle layer compositions leads to a larger range in possible BSM densities and moments of inertia for a given BSM composition. A lower degree of crystal fractionation than it was assumed in our models would hence lead to a smaller range of possible BSM densities and moments of inertia and allow stronger constraints on the BSM FeO content than we have calculated here. With our approach to assume pure fractional crystallization, we consider the maximum possible range of BSM FeO contents and thereby ensure that our estimates include as little bias as possible towards a specific scenario of lunar mantle differentiation.

# 7. Summary and conclusions

In this study we investigated the relation between the BSM FeO content and the densities and proportions of the chemical reservoirs formed by LMO solidification, in order to find an approach to better constrain BSM compositions. To this end we developed a LMO crystallization model that is consistent with recent fractional crystallization experiments, set up a simple model to simulate different mantle overturn scenarios by varying mantle stratigraphies and degrees of mixing and determined the mass distribution in the lunar interior for different overturn models to test their consistency with geophysical properties like the bulk Moon density and BSM moment of inertia. We found that:

a) Recent experiments simulating the fractional crystallization of lunar magma oceans cannot be accurately reproduced using the commonly used crystallization softwares alphaMELTS and SPICES. Both modeling softwares achieve good fits for parts of the crystallization sequence but fail to reproduce the experimental data in other parts of the sequence. Since both softwares use different modeling approaches, they have different limitations that need to be considered when applying them to model LMO solidification.

b) Recent fractional crystallization experiments simulating lunar magma ocean solidification can be reproduced best by using a combination of the crystallization softwares SPICES and alphaMELTS, where SPICES is used for the early crystallization stages until the first appearance of orthopyroxene and alphaMELTS is used for the later crystallization stages. This approach considers the strengths and weaknesses of both crystallization softwares and is applicable to lunar magma oceans with terrestrial refractory element contents, variable #Mg, and depths of ~ 600 – 1400 km.

c) The variation of BSM FeO contents systematically changes IBC layer thickness and composition. The properties of other mantle reservoirs are only slightly affected by changes in mineral chemistry and a passive reduction in reservoir volume as a result of the growing IBC layer.

d) Due to its high density, the IBC layer thickness and the efficiency of IBC sinking and mixing with the underlying mantle strongly influence the BSM moment of inertia. At high FeO contents, which induce the formation of thick IBC layers, IBC sinking can have a stronger influence on the BSM moment of inertia than the overturn of the other mantle layers.

e) Using our approach we can constrain the BSM FeO content as a function of the temperature and the degree of mixing and overturn of mantle layers. Additional information on the lunar mantle stratigraphy and local densities from seismic velocities and selenodetic data can be employed to further constrain the range of BSM FeO contents and plausible fractions of sunken IBC.

f) The assumed present-day selenotherm has a notable influence on calculated BSM densities and hence on the range of estimated BSM FeO contents. Differences in the temperature estimates of the deep lunar mantle of 450 K result in differences of ~1.5 wt% in the estimated BSM FeO content.

g) Considering all available constraints on core size and composition and mantle stratigraphy from seismic and selenodetic data as well as numerical models and petrological studies, our model favors BSM FeO contents of 9.4 – 10.9 wt% (with a lowermost limit of 9.0 wt% and an uppermost limit of 11.9 wt%) and weighted fractions of overturned IBC of 21 – 59 %. These BSM FeO contents are lower than the commonly assumed ~12 – 13 wt% FeO (e.g. Elkins-Tanton et al., 2011 and references therein), but strikingly consistent with estimates by Warren (2005) (8.3 – 10.4 wt% FeO, preferred model 9.1 wt% FeO) and suggest a moderate enrichment of the bulk silicate Moon compared to the bulk silicate Earth (~8 wt%, McDonough and Sun, 1995).

**Funding**

This work was funded by the Deutsche Forschungsgemeinschaft (DFG, German Research Foundation) – Project-ID 263649064 – TRR 170. This is TRR 170 Publication No. XX.

**CRediT author statement**

**Sabrina Schwinger:** Conceptualization, Methodology, Software, Validation, Formal analysis, Investigation, Writing - Original Draft, Visualization **Doris Breuer:** Writing - Review & Editing, Funding acquisition

**Figures and Tables**

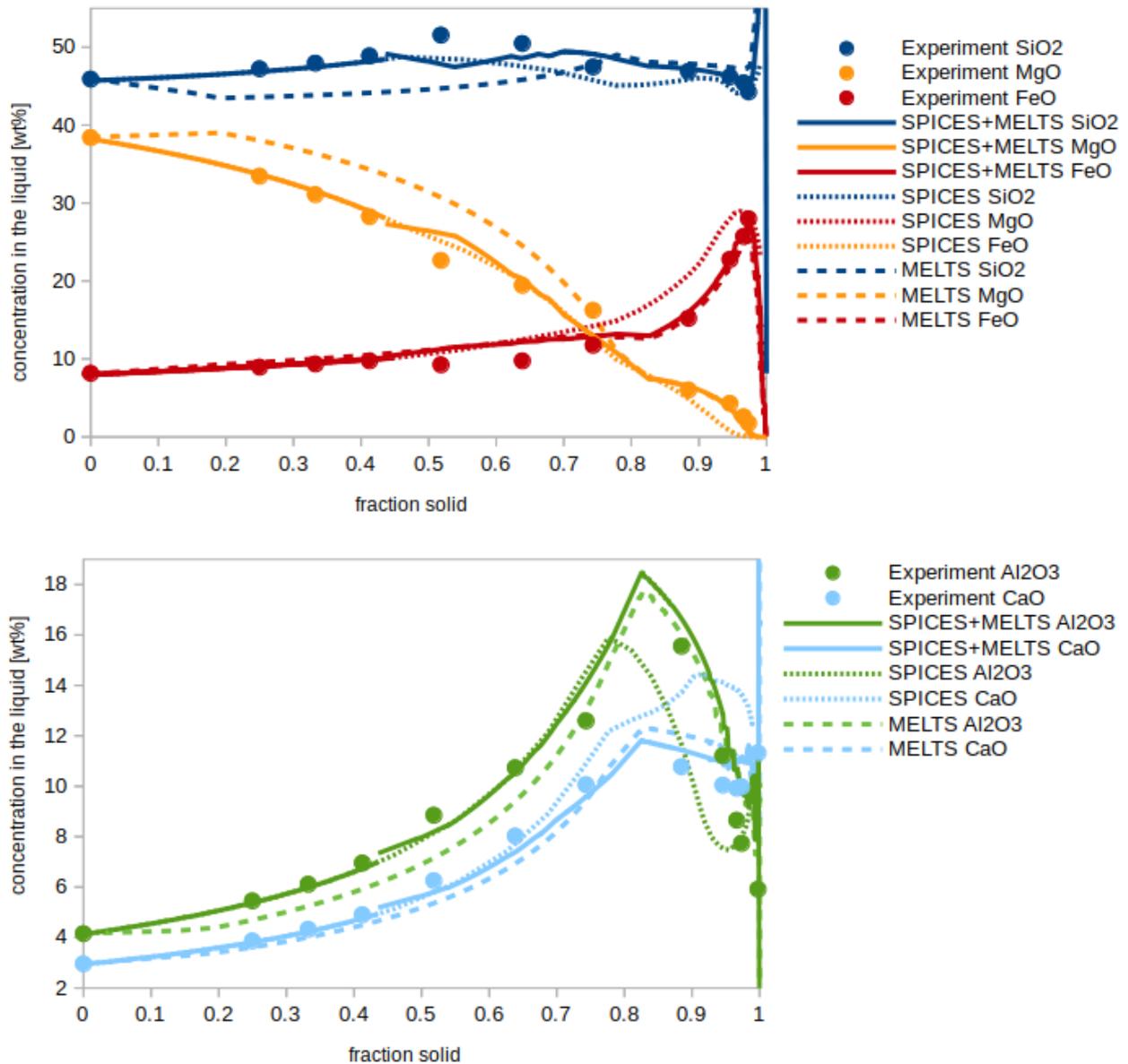

**Figure 1:** Compositional evolution of the LMO liquid during fractional crystallization. The colors represent different main oxides. Colored dots represent experimental results by Rapp and Draper (2018), colored lines are modeling results using either MELTS (dashed lines), SPICES (dotted lines) or a combination of SPICES and MELTS (solid lines).

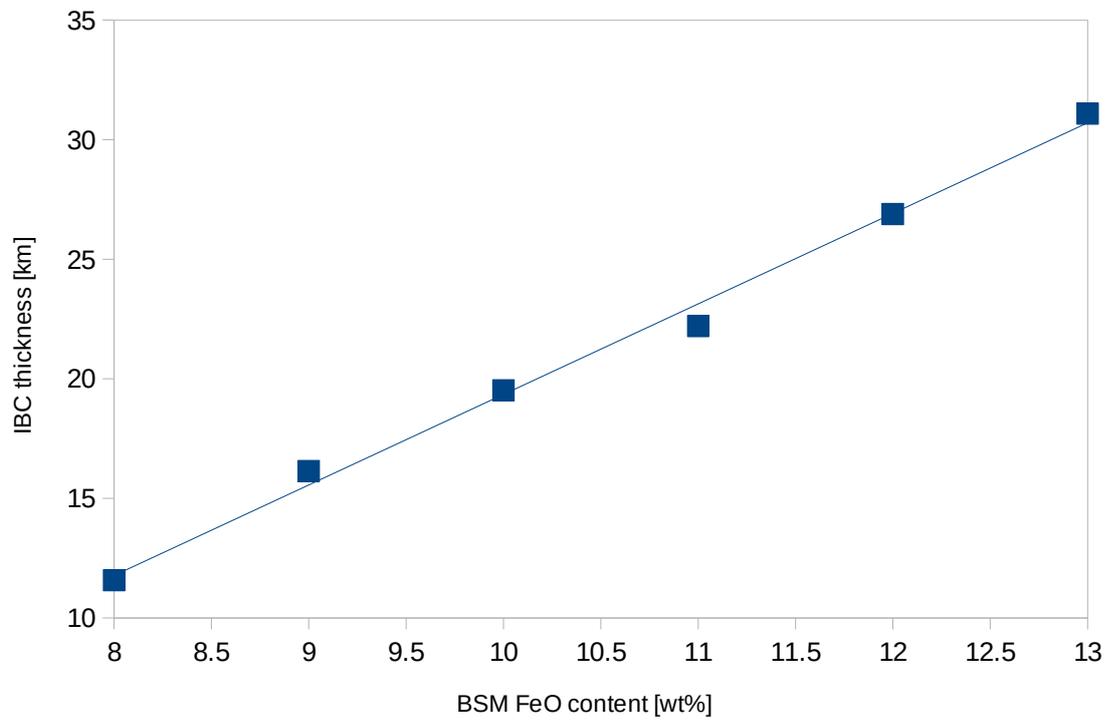

**Figure 2:** Increase of the IBC layer thickness (calculated at the position of formation) as a function of the FeO content in the LMO.

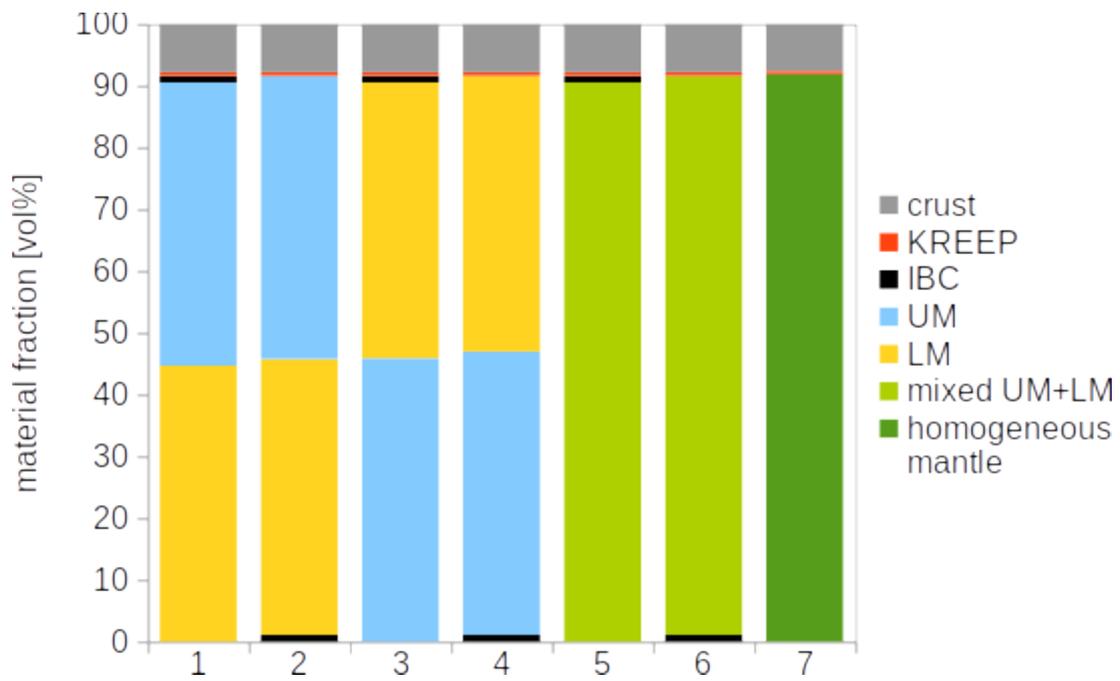

**Figure 3:** Stratigraphic models representing different overturn scenarios and degrees of mixing in the lunar interior. The construction of the compositional reservoirs is described in the text. Models 1-4 assume that the main compositional reservoirs remain separate during overturn while models 5 and 6 assume homogenization of the UM and LM reservoirs and model 7 homogenization of the UM, LM and IBC reservoirs.

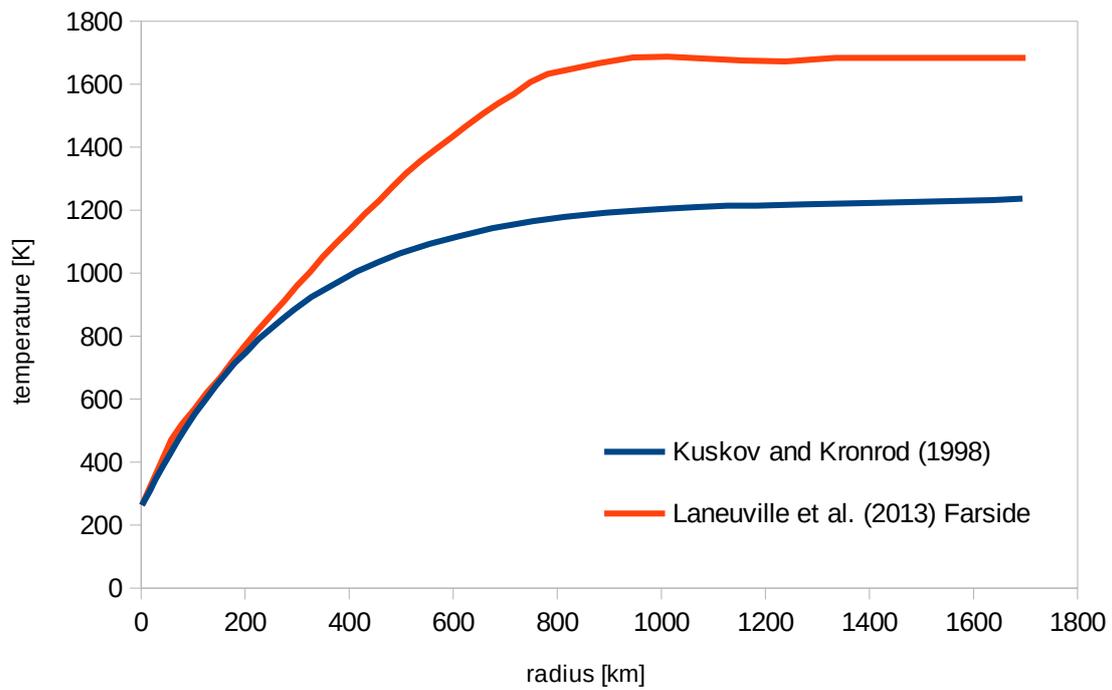

**Figure 4:** Selenotherms assumed in the lunar interior models. The hot selenotherm was chosen to allow partial melting of mantle material at the core mantle boundary (Mallik et al., 2019), while the cold selenotherm ensures that temperatures in the region of deep moonquakes are cold enough for brittle deformation (Kawamura et al., 2017).

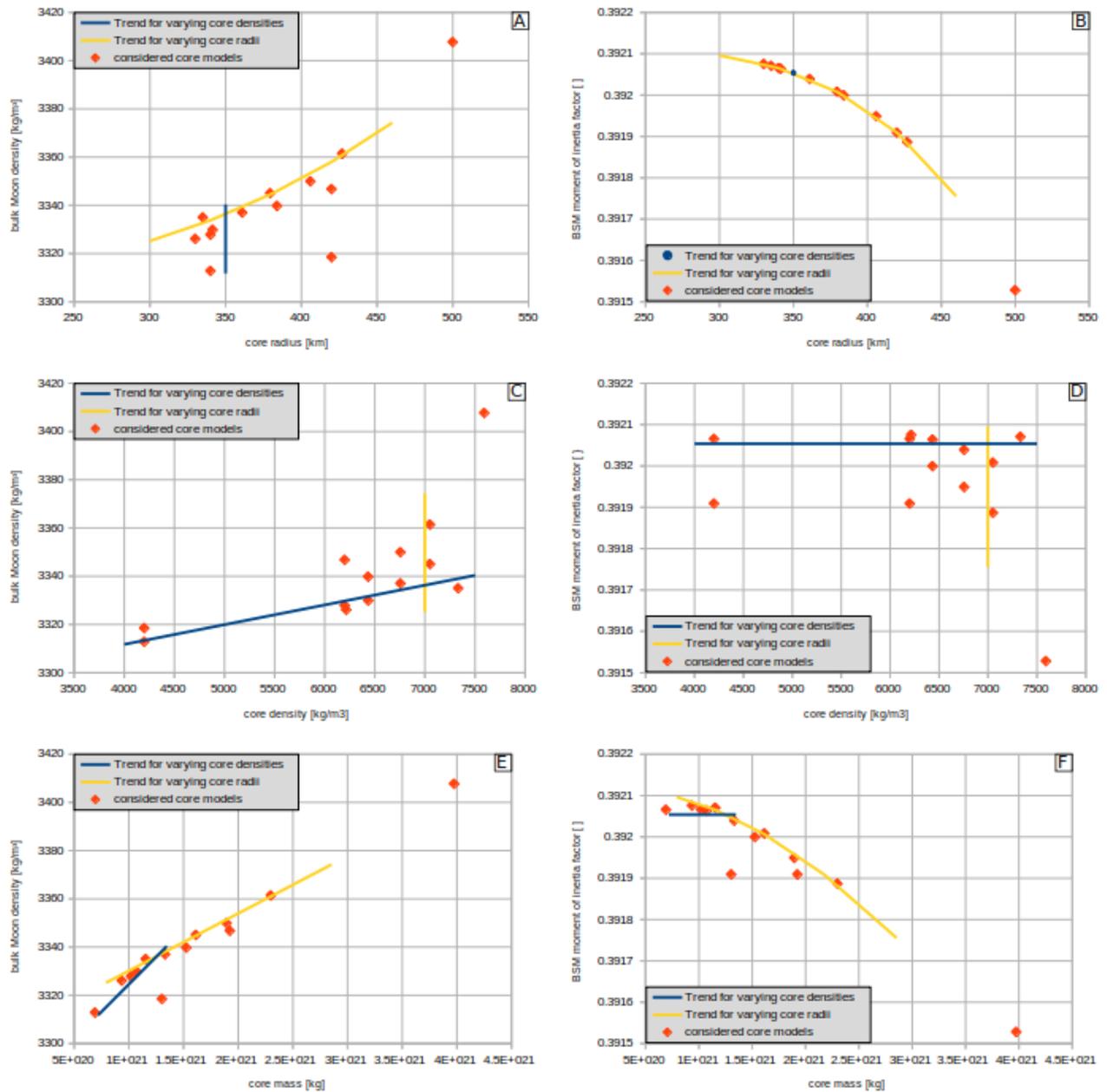

**Figure 5:** Effects of core radii, densities and masses on the bulk Moon density and the bulk silicate Moon moment of inertia factor, assuming lunar interior model 1 (see Figure 2), a BSM FeO content of 9 wt% and a cold selenotherm (see Figure 3).

The red dots represent the core models considered for calculations of possible lunar interior properties. The colored lines indicate systematic trends in the change of the bulk Moon density and the bulk silicate Moon moment of inertia factor for either varying core density for constant core size (blue) or varying core radius for a constant (yellow).

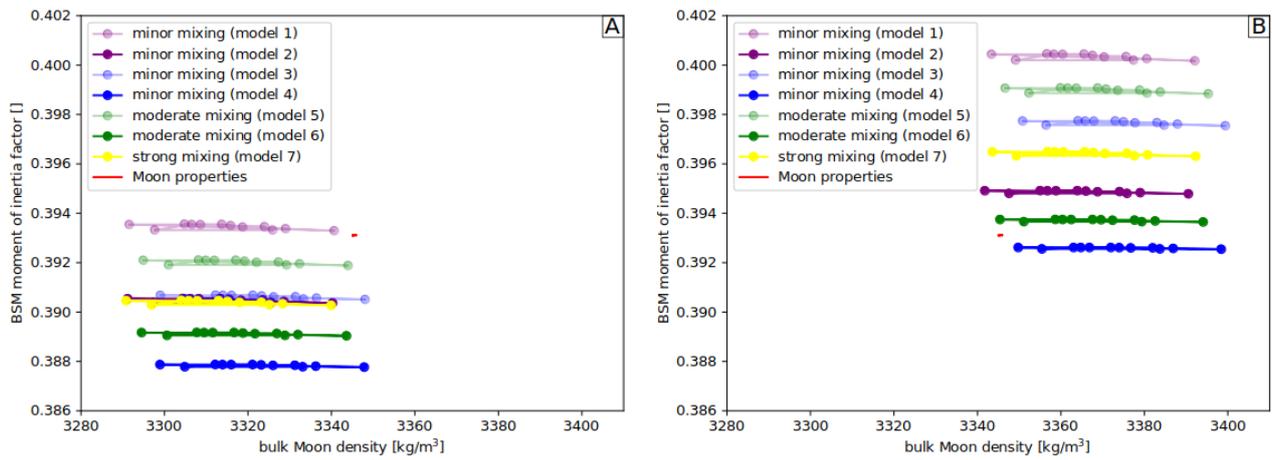

**Figure 6:** Relation of bulk Moon density and BSM moment of inertia for the different stratigraphic models described in Figure 2, assuming different BSM FeO contents (a: 9 wt%, b: 13 wt%) and a hot selenotherm (lunar Farside model, see Figure 3). The colors correspond to different degrees of mixing and overturn of the upper and lower mantle reservoirs. Light colors symbolize models without IBC overturn, dark colors represent models with full IBC overturn. The red square represents the measured bulk Moon density and BSM moment of inertia with uncertainties as reported by Matsumoto et al. (2015).

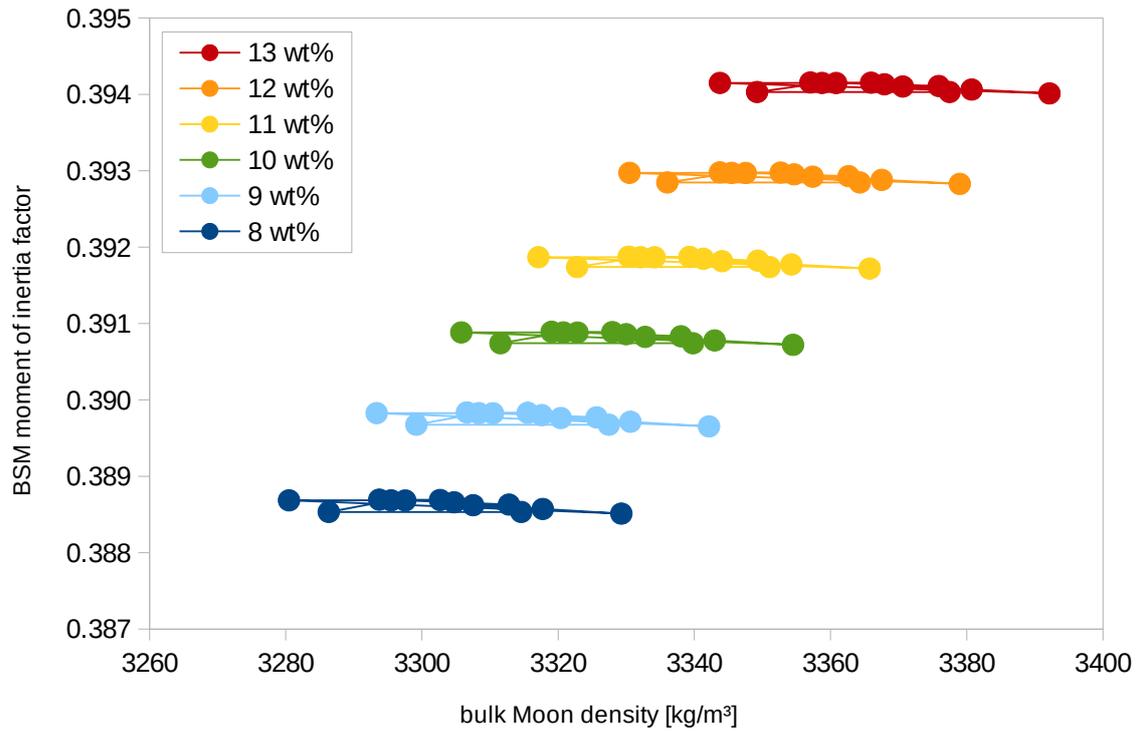

**Figure 7:** Relation of bulk Moon density and BSM moment of inertia for different BSM FeO contents, assuming lunar interior structure model 1 (see Figure 2) with various core models (represented by individual dots) and a cold selenotherm after Kuskov and Kronrod (1998), (see Figure 3).

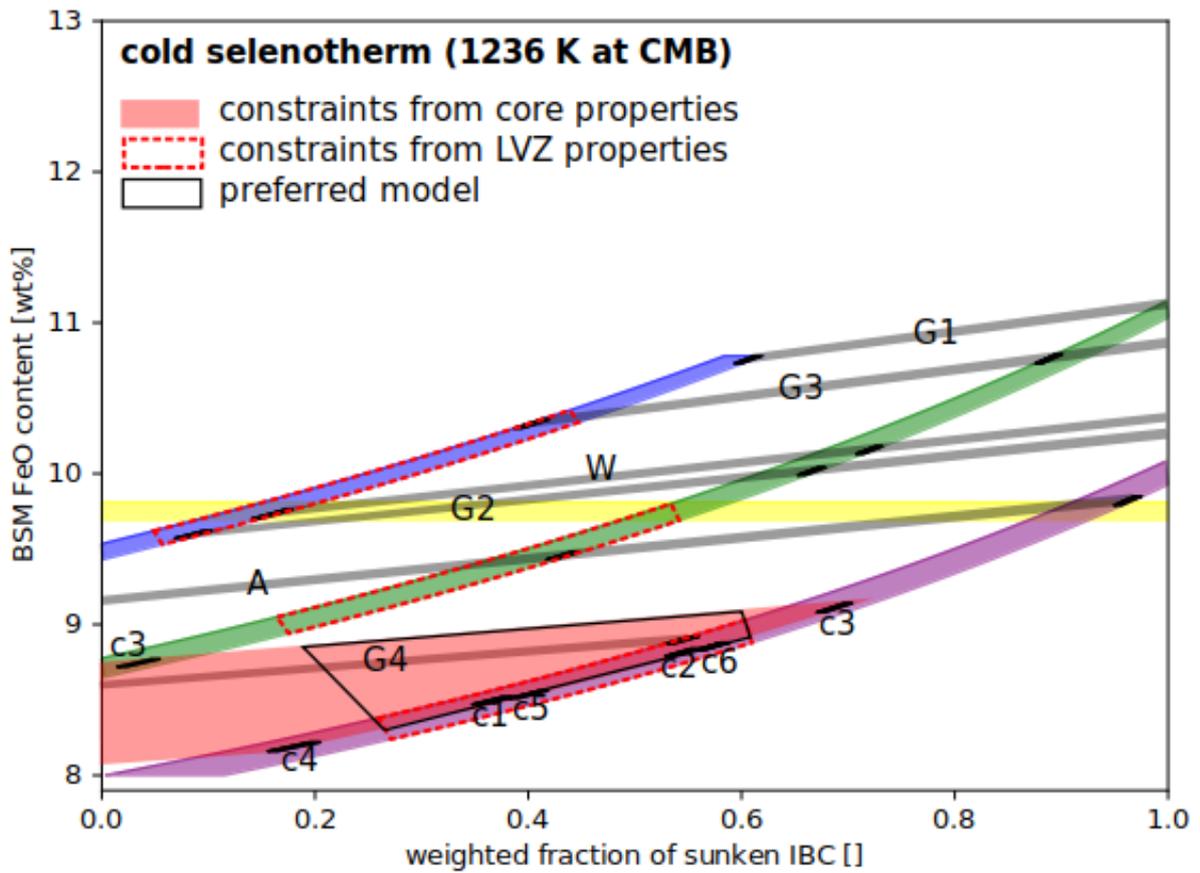
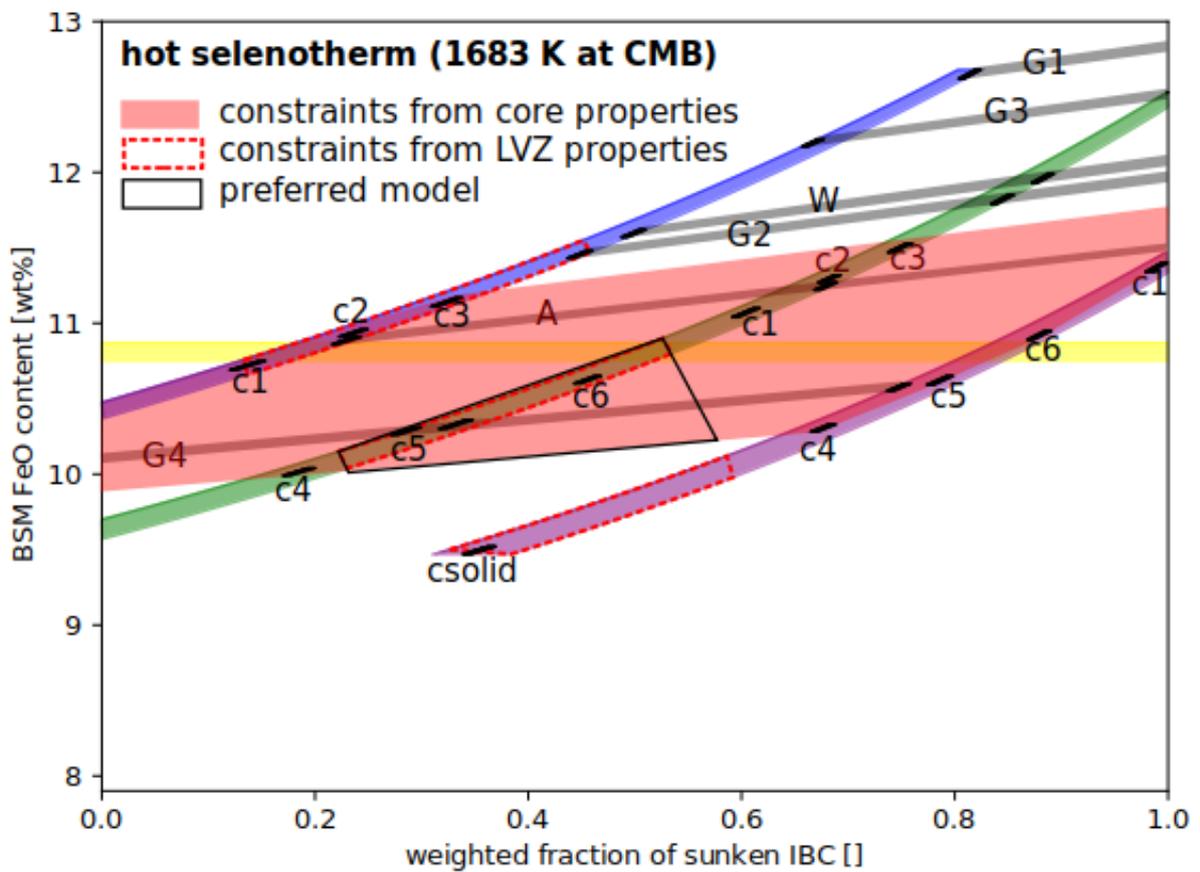

**Figure 8:** Relation of BSM FeO content and fraction of overturned IBC material for different mixing and overturn scenarios, assuming the different selenotherms shown in Figure 4. A: cold selenotherm after Kuskov and Kronrod (1998), B: hot selenotherm proposed by Laneuville et al. (2013) for the Lunar Farside. Note that the color code for the interior structure models is the same as in Figure 6. The parts of the parameter space that are also consistent with the low velocity zone density and thickness proposed by Matsumoto et al. (2015) are bordered by red dotted lines.

The black areas connected by gray lines represent the parameter space that was calculated for specific core models from the literature (G1-G4: Garcia et al., 2011; W: Weber et al., 2011; A: Antonangeli et al., 2015). The properties of these cores and associated bulk Moon models are listed in table 2. Unconnected black areas represent core models (c1-c6) calculated for core sizes consistent with recent LLR data (Viswanathan et al. 2019) (380±12 km) and densities calculated for a maximum core S content of 8 wt% and the CMB temperatures of the respective selenotherms. The red area marks the parameter space consistent with these core models.

Black lines encircle the parameter space that is consistent with all considered constraints and can hence be considered as most realistic for the Moon.

| Parameter | Values | References |
|---|---|---|
| LMO depth | ~1350 km (full BSM melting) | Rai and van Westrenen (2014) |
| BSM FeO contents | 8 – 17 wt% | Jones and Delano (1989), Warren (1986) |
| Crustal thickness | 40 km | Wieczorek et al. (2013) |
| Crust density | 2900 kg/m³ | Huang and Wieczorek (2012) |
| Core radii | 255 – 427 km | Garcia et al. (2011), Weber et al. (2011), Antonangeli et al. (2015) |
| Core densities | 4200 – 7333 kg/m³ | Garcia et al. (2011), Weber et al. (2011), Antonangeli et al. (2015) |
| Mean lunar radius (R) | 1737.151 km | Matsumoto et al. (2015) |
| Bulk lunar mass (M) | $(7.34630 \pm 0.00088) \times 10^{22}$ kg | Matsumoto et al. (2015) |
| Bulk Moon density | 3345.54855 ± 0.40075 | Matsumoto et al. (2015) |
| Normalized BSM moment of inertia (I/MR²) | 0.393112 ± 0.000012 | Matsumoto et al. (2015) |

**Table 1:** Parameters used for the construction of lunar interior models.

| Core model / Mantle model | Core density [kg/m³] | Core radius [km] | Bulk silicate Moon FeO [wt%] | | Weighted fraction of sunken IBC [] | |
|---|---|---|---|---|---|---|
| | | | Cold selenotherm | Hot selenotherm | Cold selenotherm | Hot selenotherm |
| W (Weber et al., 2011) | 6216 | 330 | 9.7 – 10.4 | 11.6 – 12.1 | 0.14 – 1.0 | 0.49 – 1.0 |
| A (Antonangeli et al., 2015) | 7332 | 335 | 9.1 – 9.8 | 10.9 – 11.5 | 0.0 – 0.98 | 0.21 – 1.0 |
| G1 (Garcia et al., 2011) | 4200 | 340 | 10.7 – 11.1 | 12.6 – 12.9 | 0.6 – 1.0 | 0.8 – 1.0 |
| G2 (Garcia et al., 2011) | 6200 | 340 | 9.6 – 10.3 | 11.4 – 12.0 | 0.07 – 1.0 | 0.44 – 1.0 |
| G3 (Garcia et al., 2011) | 4200 | 420 | 10.3 – 10.9 | 12.2 – 12.6 | 0.39 – 1.0 | 0.66 – 1.0 |
| G4 (Garcia et al., 2011) | 6200 | 420 | 8.5 – 8.9 | 10.1 – 10.6 | 0.0 – 0.56 | 0.0 – 0.76 |
| Purple, with c1-6 (with c1-6 and LVZ) | 6860 – 7666 | 381±12 | 8.1 – 9.2 (8.2 – 9.1) | 10.2 – 11.5 (n.a.) | 0.07 – 0.72 (0.26 – 0.61) | 0.63 – 1.0 (n.a.) |
| Blue with c1-6 (with c1-6 and LVZ) | 6860 – 7666 | 381±12 | n.a. (n.a.) | 10.4 – 11.2 (10.7 – 11.2) | n.a. (n.a.) | 0.0 – 0.37 (0.13 – 0.37) |
| Green with c1-6 (with c1-6 and LVZ) | 6860 – 7666 | 381±12 | 8.6 – 8.8 (n.a.) | 10.0 – 11.6 (10.0 – 10.9) | 0.0 – 0.1 (n.a.) | 0.12 – 0.78 (0.22 – 0.53) |
| Yellow | not specified | not specified | 9.7 – 9.8 | 10.8 – 10.9 | n.a. | n.a. |
| Preferred model (preferred model with LVZ) | 6860 – 7666 | 381±12 | 8.1 – 9.2 (8.3 – 9.1) | 9.9 – 11.8 (10.0 – 10.9) | 0.0 – 0.73 (0.19 – 0.61) | 0.0 – 1.0 (0.22 – 0.58) |
| Preferred model with LVZ and outer core thickness 80-85 km | 6787 – 7380 | 381±12 | 9.4 – 10.4 (1525 – 1560 K, 8 wt% S in the core) | | 0.21 – 0.59 (1525 – 1560 K, 8 wt% S in the core) | |

**Table 2:** Ranges of model results (BSM FeO and weighted fraction of sunken IBC) associated with selected core models and mantle models. Densities and radii of core models W and A were calculated assuming the maximum values proposed by Weber et al. (2011) and Antonangeli et al., (2015), respectively. The densities and radii of core models G1-4 are based on the maximum and minimum values proposed by Garcia et al. (2011). The mantle models are named according to the color by which they are represented in Figure 6 and Figure 8, with purple: minor mixing (mantle not overturned), blue: minor mixing (mantle overturned), green: moderate mixing and yellow: strong mixing. For the mantle models we list both the complete range of values (corresponding to colored areas in Figure 8) and the range of values consistent with a low velocity zone as proposed by Matsumoto et al. (2015) (marked by the red dotted lines in Figure 8).

The preferred model includes only interior models consistent with core properties of c1-6, the low velocity zone properties by Matsumoto et al. (2015) and a mantle with limited overturn (models purple or green) as suggested by seismic models. To fit the outer core thickness proposed by Weber et al. (2011), we assume a different CMB temperature range but keep the same core composition.

**Supplementary Material**

**S.1. LMO solidification model**

In order to find a theoretical modeling approach that is suitable for the conditions of lunar magma ocean crystallization, we compared the results of two different crystallization models (FXMOTR and MELTS/pMELTS) with the experimental results reported by Rapp and Draper (2018). The initial compositions and experimental conditions used by Rapp and Draper (2018) are listed in Table S1.

FXMOTR is part of SPICEs, a suite of crystallization programs in a Matlab environment that has been developed by Davenport et al. (2013). It uses a combination of fractional and equilibrium crystallization and calculates phase abundances and compositions based on a set of algorithms and experimentally determined liquidus boundaries in various subprojections in the olivine-plagioclase-wollastonite-silica system. The MAGFOX and MAGPOX algorithms for fractional and equilibrium crystallization included in SPICEs have originally been developed as Fortran programs (e.g. Longhi, 1980, 1982) and have been applied for the modeling of magmas on various planetary bodies (e.g. Neal et al., 1994; Brown and Elkins-Tanton, 2009). Most notably they were applied by Snyder et al. (1992) to calculate a LMO crystallization sequence that is commonly used in studies of lunar mantle evolution (e.g. Elkins-Tanton et al., 2011).

The MELTS and pMELTS algorithms calculate phase properties by Gibbs energy minimization using a database of experimentally determined thermodynamic properties of minerals and silicate melt. pMELTS is specifically calibrated for peridotite compositions at elevated pressures (1-3 GPa), while MELTS covers a larger range of compositions and is most reliable at lower pressures (0-2 GPa). MELTS and pMELTS have been widely used to study terrestrial magmatic systems but has also been applied to study extraterrestrial magmatic rocks (e.g. Slater et al., 2003; Thompson et al., 2003).

In order to test the consistency of the crystallization models with experimental results, we attempted to fit the results of crystallization experiments by Rapp and Draper (2018) with FXMOTR, MELTS and pMELTS. Thereby we attempted to reproduce both individual crystallization experiments, assuming the conditions used by Rapp and Draper (2018) (see Table S1), and the complete crystallization sequence, starting from the initial pressure and composition used by Rapp and Draper (2018) in their first crystallization step. In addition, we tested different combinations of crystallization algorithms in order to find a model that fits the experimental data in terms of mineral modal abundances and degrees of solidification for given pressures, temperatures and compositions.

As it has been noted by Ghiorso et al. (2002), the pMELTS algorithm overestimates the stability of garnet at high pressures. According to experimental studies, e.g. by Elardo et al. (2011), garnet is not a liquidus phase in the lunar magma ocean and is hence unlikely to form in a fractionally crystallizing magma ocean. Therefore garnet crystallization was suppressed in the MELTS/pMELTS crystallization model. FXMOTR prescribes that olivine is the liquidus phase at the beginning of magma crystallization, so that early garnet crystallization is excluded by default.

### *S.1.1. Reproduction of individual experiments*

The temperatures and mineralogies of individual experimental results from Rapp and Draper (2018) and the respective results for different crystallization models are shown in figures S1 and S2. As shown in figure S2, both the pMELTS and the MELTS algorithms fail to reproduce the correct mineral modal abundances of the first four experiments (LPUM – L-PC5) and experiment L-PC14 by overestimating the stability of pyroxene compared to olivine and plagioclase. Accordingly, the crystallization temperatures predicted by MELTS are systematically overestimated by 125-265 K in the early crystallization stages (figure S1). For the late stage crystallization experiments temperatures are only overestimated by a few 10s of K.

In the last crystallization experiment (L-GM24) the degree of solidification is very high (85%) and the run products contain both olivine and quartz, which is a thermodynamically unstable phase assemblage and indicates that the experimental sample did not reach thermodynamic equilibrium. Indeed the MELTS equilibrium crystallization model predicts olivine to crystallize first but to completely react with the melt to clinopyroxene during further cooling of the system. This reaction might have been not fully completed in the experiment, which explains why the experimental run contains both olivine and quartz, while the equilibrium model calculation does not. In addition it explains why the experiment is the only one for which the MELTS model *underestimates* the solidus temperature instead of overestimating it (because it assumes a higher content of low temperature phases).

The presence of apatite in the last experiment (L-GM24) indicates that some of the nominally dry experimental charges were contaminated with water. Water is known to lower the solidus of silicate melts, so not considering water in a melt generally leads to an overestimation of crystallization temperatures. To test to which degree the effect of water can explain the observed temperature deviations, we added some water to the compositions and compared the calculated crystallization temperatures for the dry and wet model. At water saturation the modeled crystallization temperature for the experimentally observed degree of crystallization was about 11K lower than the respective crystallization temperature calculated at dry conditions. This temperature difference can be interpreted as the maximum possible temperature deviation induced by the presence of water. So the putative presence of water in the experiments can explain only a part of the deviation between modeled and observed crystallization temperatures.

The pMELTS algorithm fails to reproduce the modal abundances of experiments L-PC20 and L2-GM21 (mainly by overestimating the stability of olivine) and did not find a solution for the L-GM24 composition. These three experiments were conducted at 1 bar and hence lie outside of the recommended pressure range for pMELTS (1-3 GPa).

FXMOTR reproduces the mineralogies of the first six experiments (LPUM - L-PC10) perfectly (figure S2) and also is best at reproducing the crystallization temperatures in this part of the crystallization sequence (figure S1). However, in the remaining experiments the algorithm either fails to find a result or severely overestimates the stability of plagioclase. This indicates that FXMOTR is only applicable for compositions corresponding to the first crystallization stages. Since FXMOTR always assumes olivine to be the liquidus phase, it is not suited for the calculation of individual experiments with bulk compositions for which olivine is not on the liquidus. However, FXMOTR is able to model fractional crystallization sequences of initial compositions that involve early olivine crystallization followed by late stages without olivine crystallization. Therefore we also calculated complete crystallization sequences with MELTS, pMELTS and FXMOTR starting with the LPUM composition to compare the general crystallization trends with the experimental results.

### S.1.2. General crystallization trends

For the complete crystallization sequences we assumed LPUM starting compositions and initial pressures of 4 GPa. Pressures were adjusted according to the degree of crystallization after every crystallization step. FXMOTR calculations were performed at steps of 0.1 PCS (percent solid) and MELTS and pMELTS calculations at temperature steps of 1K. To account for the recommended conditions of application for MELTS and pMELTS, both algorithms were integrated in a single crystallization model, using pMELTS at higher pressures (> 0.7 GPa) and MELTS at lower pressures.

The crystallization temperatures, the evolution of mineral modal abundances during crystallization and the final mineral modal abundances after crystallization are shown in figures S3, S4 and S5.

Figure S6 illustrates the compositional evolution of the melt during fractional crystallization for different computational models and the experiments by Rapp and Draper (2018). The pMELTS/MELTS model systematically underestimates liquid $SiO_2$ contents and overestimates liquid MgO contents, which corresponds to the erroneous prediction of pyroxene crystallization instead of olivine in the early crystallization stages (figure S2 and S4). FeO, $Al_2O_3$, CaO and $TiO_2$ contents are reproduced well with deviations of typically less than 3 wt% from the experimental values. In the experiment 50% of plagioclase crystallization is completed at 88 PCS, while in the model plagioclase formation is either slightly delayed or proceeds slower, so that 50% of the plagioclase has formed only at 90 PCS (figure S4). The total plagioclase content is about 10% lower than in the experimental series. The MELTS algorithm underestimates the amount of ilmenite and overestimates the amount of other Ti bearing phases like spinel and sphene. However, the total amount of Ti bearing phases is the same as the ilmenite content in the experiment (figure S5) and the formation of Ti-bearing phases occurs at the same time in the crystallization sequence (at 97 PCS). This indicates that the MELTS algorithm correctly represents the oversaturation of Ti in the melt (see figure S6) even though it does not calculate the correct mineralogy. Sphene ($CaTiSiO_5$) crystallization removes additional $SiO_2$ from the melt which is reflected in a lower total quartz content compared to the experiment.

The FXMOTR model reproduces the early compositional evolution (<75 pcs) of the liquid for all elements shown in figure S6 as well as the mineralogy of the cumulate (figure S4). However, during the later crystallization stages the liquid $Al_2O_3$ and CaO contents are underestimated (figure S6), corresponding to an overestimation of the amount of plagioclase (figure S4). In addition late liquid FeO contents are overestimated and $TiO_2$ contents underestimated (figure S6). The latter is related to an overestimation of the amount of crystallizing ilmenite (figure S4).

### S.1.3. A combined modeling approach

Neither pMELTS/MELTS nor FXMOTR were successful in reproducing the compositional evolution of the liquid (figure S6) and the mineral modal abundances in the cumulate (figure S4) due to their specific limitations. However, both models complement each other in that FXMOTR succeeded in reproducing the early crystallization history, while MELTS/pMELTS produced accurate results in the late crystallization stages. Therefore we tested a combined modeling approach where FXMOTR was used for early and MELTS for later steps of fractional crystallization. The degree of solidification at which we switched from FXMOTR to MELTS was varied to find the best fit to the bulk abundances of minerals, the temperatures of their first appearance and the chemical evolution of the melt.

The main problem of the pMELTS algorithm in predicting the correct mineralogies lies in the underestimation of olivine stability at high pressures (Ghiorso et al. 2002), that leads to crystallization of orthopyroxene before olivine in deep LMO settings. FXMOTR on the other hand correctly predicts the transition from olivine to orthopyroxene crystallization in the early stages of LMO solidification. Hence the best fit for the crystallization sequence is achieved by a combined model in which the early stages of crystallization (up to ~ 45 pcs, just before Opx becomes stable) are calculated with FXMOTR, while the later stages are modeled with pMELTS/MELTS. This approach also produces the best fits for the compositional evolution of the melt, crystallization temperatures and bulk mineralogy (see figures S3, S4, S5 and S6).

### *S.1.4. Verification with additional experimental data*

To test the applicability of this modeling approach to other compositions and magma ocean depths, we fitted experimental results by Charlier et al. (2018) for an FeO-rich LMO composition proposed by O'Neill (1991), with a low MgO/(MgO+FeO) ratio of only 0.74 compared to 0.81 for the LPUM composition by Rapp and Draper (2018). Charlier et al. (2018) assumed shallow magma ocean depths of 600 km and performed fractional crystallization experiments for late stages of magma ocean crystallization, using starting compositions derived by assuming initial olivine precipitation from the bulk LMO composition.

The evolution of the cumulate mineralogies during crystallization (figure S7) as well as the crystallization temperatures (figure S8) are well reproduced by the crystallization model, which indicates that our modeling approach is applicable to magma ocean compositions with variable #Mg and terrestrial refractory element contents. Charlier et al. (2018) also performed fractional crystallization experiments for a TWM composition that is enriched in $Al_2O_3$ and CaO compared to the Earth's mantle. However, current estimates of the crust thickness, modeling of seismic data and the mantle source compositions of mare basalts indicate that the $Al_2O_3$ content of the lunar mantle is not higher than about 3.5 wt% (Taylor et al. 2013 and references therein). Therefore we assume that the TWM composition is unlikely to reflect lunar mantle compositions and did not consider it for evaluating our modeling approach.

**S.2. Core models**

To obtain a set of realistic core models, we calculated pausible core masses and densities consistent with recent studies on lunar core size and composition. We calculated two sets of core models consistent with core temperatures of 1400 K and 1600 K, respectively, corresponding to the two different selenotherms assumed in our lunar interior models. Each set contains 6 core models that differ in their radius (3 different radii) and density (two different densities). The variations in outer core radius correspond to minimum, maximum and average values of the core radius estimate of 381 ± 12 km by Viswanathan et al. (2019). Estimates of the maximum and minimum core density were made considering the effects of temperature and core composition on the degree of core melting and the density of the solid and liquid core. The properties of all core models are listed in Table S2.

The degree of melting of the core was determined using the phase diagrams reported by Liu and Li (2020), assuming a homogeneous core temperature equal to the temperature at the core mantle boundary, a bulk core S content of 8 wt% and bulk core Ni contents of 0-46 mol%. The same phase diagrams were used to estimate the S content of the liquid outer core in each model.

The assumed inner core density corresponds to the density of pure iron at the assumed core temperature. The density of the liquid outer core was determined based on the S content of the outer core and the density estimates by Antonangeli et al. (2015) for Fe-S liquids. Since Antonangeli et al. (2015) report density data for ~1800 K, we applied a correction to consider different core temperatures, assuming that the Fe-S liquid has a similar volumetric thermal expansion coefficient as solid iron (Blumm and Henderson, 2000). Due to the large uncertainty in the density estimates, we determined upper and lower values for each core temperature.

Different models for the position of the inner core boundary were combined with the density estimates in such a way as to produce models of maximum and minimum bulk core density (i.e. combining the models with maximum degree of melting with the minimum estimates of liquid density and combining the models with minimum degree of melting with the maximum estimates of liquid density).

**Figures**

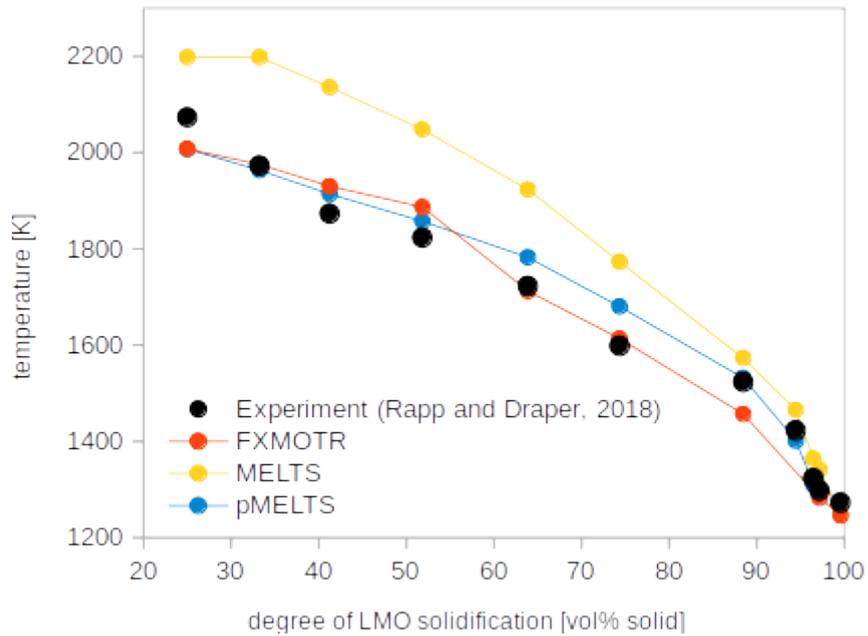

**Figure S1:** LMO temperature as a function of the degree of LMO solidification. Black dots represent the temperatures of single experiments by Rapp and Draper (2018), while the colored dots indicate temperatures calculated with different phase equilibria programs for the same compositions and degree of solidification as in the respective experiments.

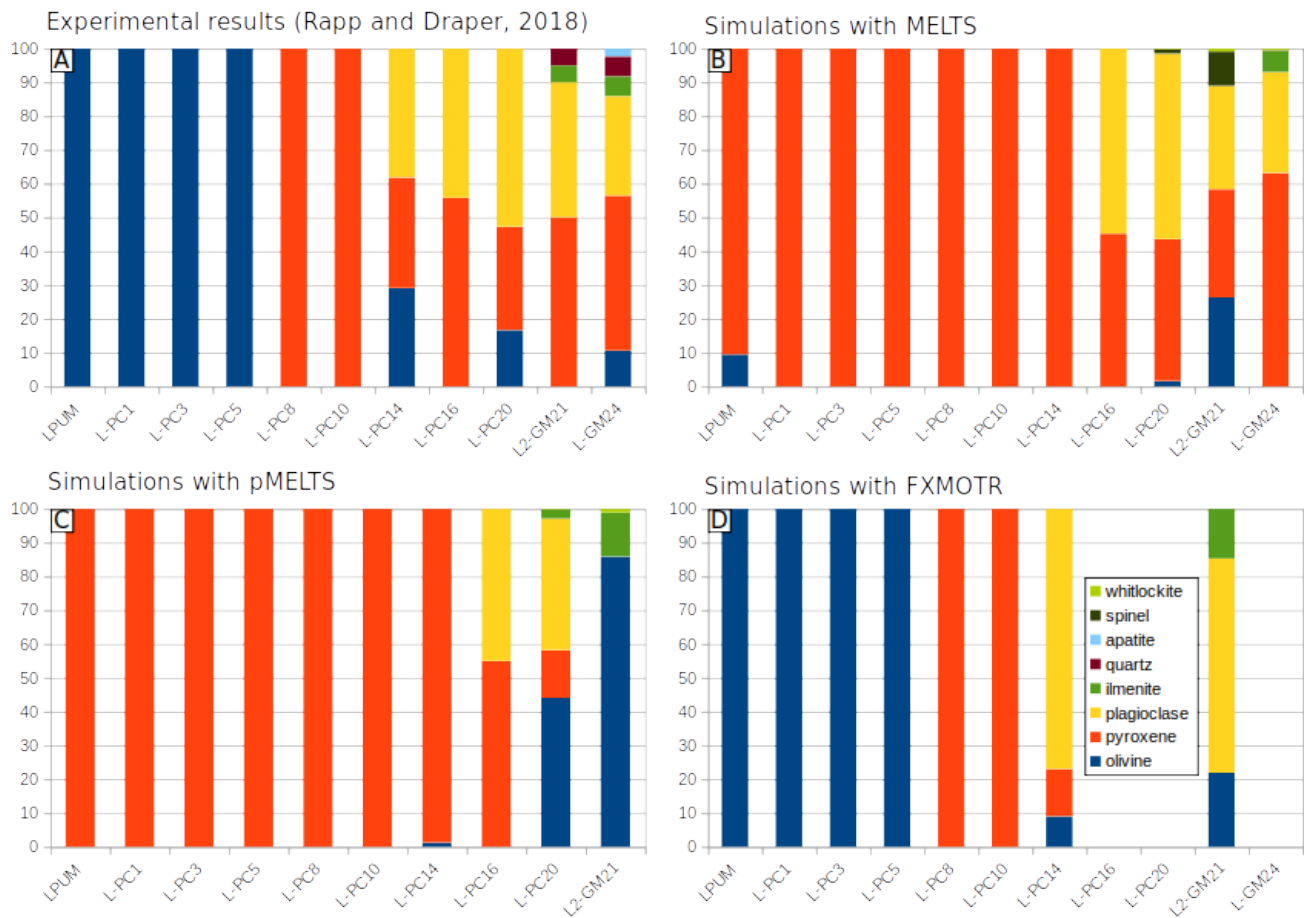

**Figure S2:** Mineral abundances in the solid fraction of single experiments by Rapp and Draper (2018) and calculated mineral abundances for the same compositions and degrees of solidification as in the respective experiments. Since the type of pyroxene is not specified in some experiments, simulated orthopyroxene and clinopyroxene fractions are combined and labeled "pyroxene" for a better comparability of experiments and simulations.

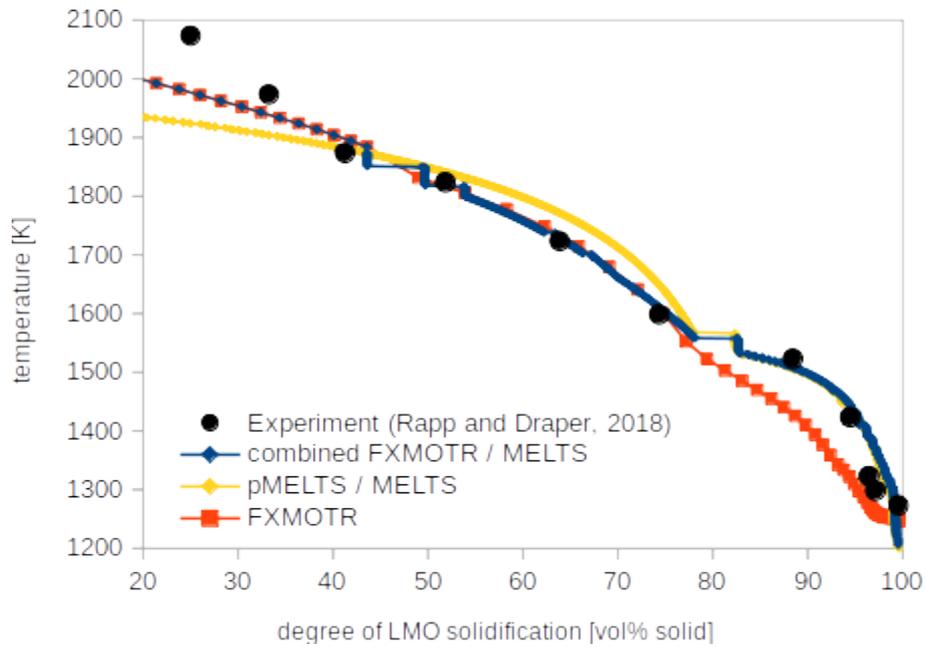

**Figure S3:** LMO temperature as a function of the degree of LMO solidification. Black dots represent the temperatures of single experiments by Rapp and Draper (2018), while colored lines and symbols indicate the LMO temperature evolution calculated with different phase equilibria programs. All calculations were made assuming fractional crystallization of the LPUM initial composition used by Rapp and Draper (2018). The crystallization models are described in detail in the main text.

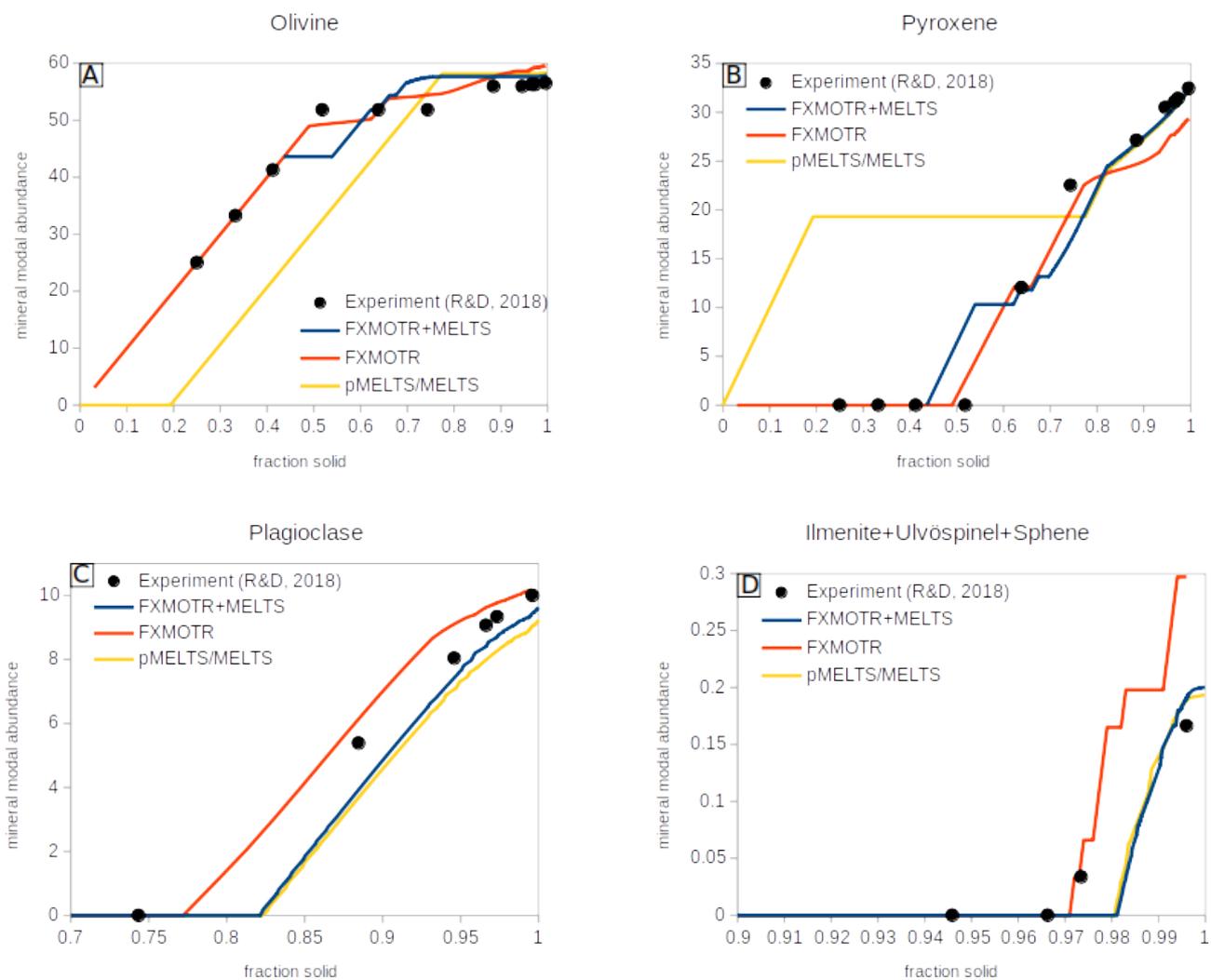

**Figure S4:** Mineral abundances as a function of the degree of solidification. All calculations were made assuming fractional crystallization of the LPUM initial composition used by Rapp and Draper (2018). The crystallization models are described in detail in the main text.

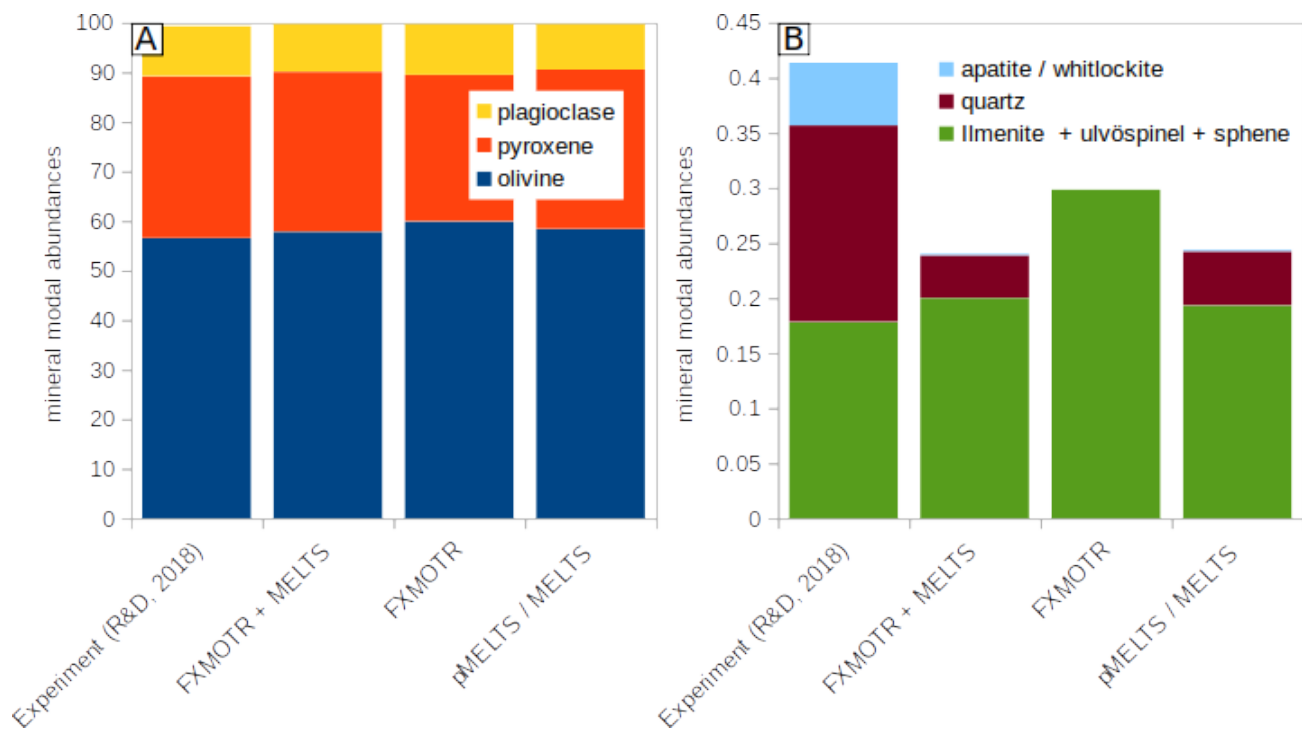

**Figure S5:** Bulk cumulate mineralogies for different crystallization models compared with experimental results by Rapp and Draper (2018). All calculations were made assuming fractional crystallization of the LPUM initial composition used by Rapp and Draper (2018). The crystallization models are described in detail in the main text.

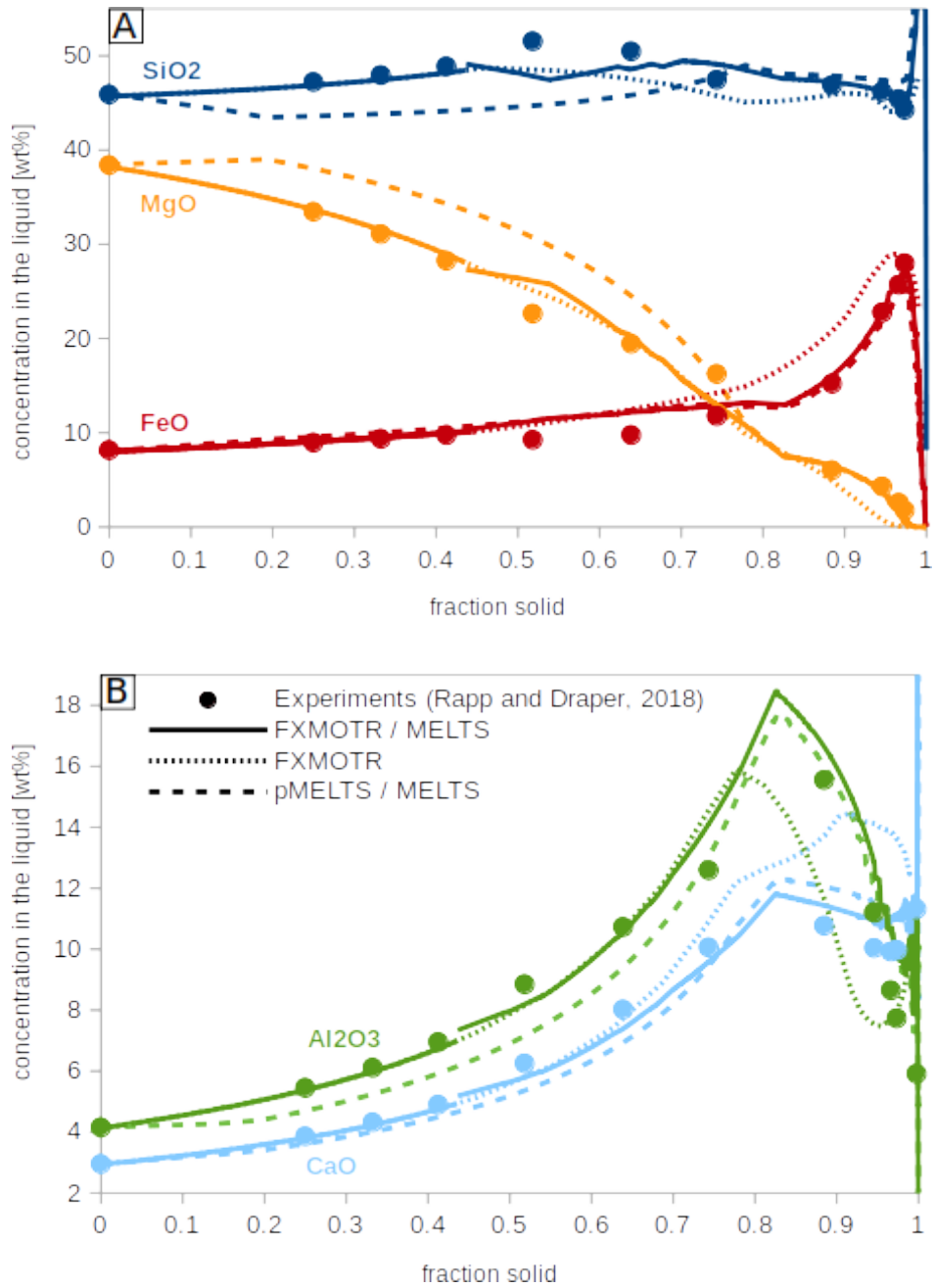

**Figure S6:** Compositional evolution of the liquid for different crystallization models compared with experimental results by Rapp and Draper (2018). All calculations were made assuming fractional crystallization of the LPUM initial composition used by Rapp and Draper (2018). The crystallization models are described in detail in the main text.

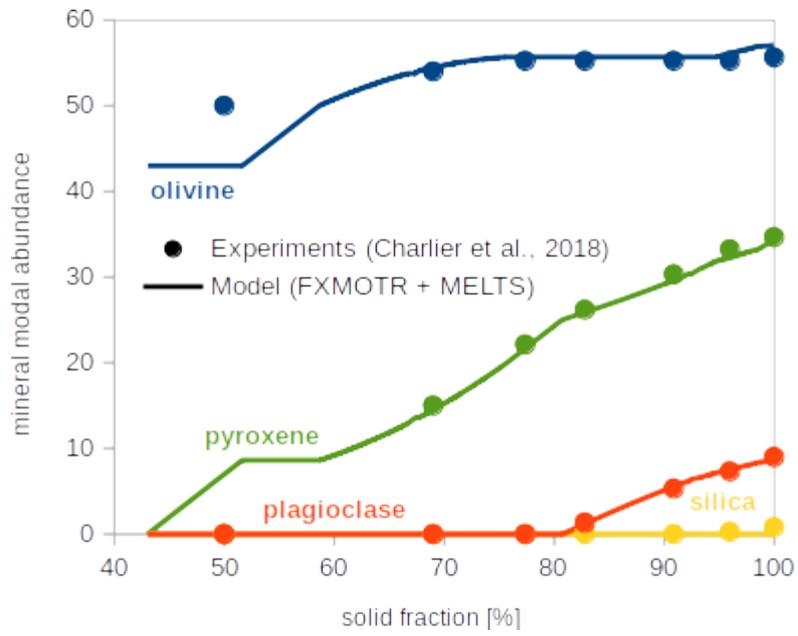

**Figure S7:** Mineral abundances as a function of the degree of solidification. All calculations were made assuming fractional crystallization of the O'Neill composition (based on O'Neill, 1991) used by Charlier et al. (2018).

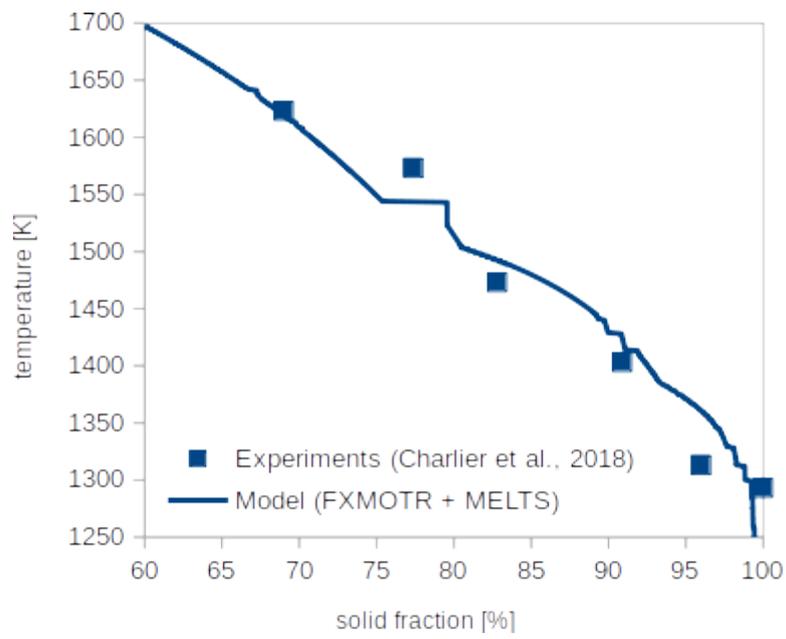

**Figure S8:** Crystallization temperatures as function of the degree of solidification. All calculations were made assuming fractional crystallization of the O'Neill composition (based on O'Neill, 1991) used by Charlier et al. (2018).

**Tables**

|       | LPUM  | L-PC1 | L-PC3 | L-PC5 | L-PC8 | L-PC10 | L-PC14 | L-PC16 | L-PC20 | L2-GM21 | L-GM24 |
|-------|-------|-------|-------|-------|-------|--------|--------|--------|--------|---------|--------|
| SiO2  | 45.9  | 47.22 | 47.95 | 48.87 | 51.54 | 50.47  | 47.45  | 46.88  | 46.18  | 45.42   | 44.26  |
| TiO2  | 0.15  | 0.2   | 0.22  | 0.25  | 0.33  | 0.41   | 0.51   | 1.33   | 2.51   | 3.7     | 4.02   |
| Al2O3 | 4.15  | 5.45  | 6.11  | 6.95  | 8.85  | 10.73  | 12.59  | 15.55  | 11.21  | 8.65    | 7.73   |
| FeO   | 8.15  | 8.94  | 9.35  | 9.75  | 9.24  | 9.75   | 11.75  | 15.2   | 22.76  | 25.7    | 27.99  |
| MnO   | 0.12  | 0.13  | 0.14  | 0.14  | 0.14  | 0.11   | 0.02   | 0.26   | 0.36   | 0.42    | 0.55   |
| MgO   | 38.4  | 33.45 | 31.08 | 28.26 | 22.64 | 19.43  | **16.23** | **6.02** | 4.29 | 2.57    | 1.76   |
| CaO   | 2.95  | 3.86  | 4.32  | 4.9   | 6.25  | 8.02   | 10.06  | 10.77  | 10.04  | 9.92    | 9.98   |

**Table S1:** Starting compositions, pressures and temperatures of the fractional crystallization experiments by Rapp and Draper (2018).

| Model | T [K] | Inner core radius [km] | Outer core radius [km] | Inner core density [kg/m³] | Outer core density [kg/m³] | Outer core S content [wt%] |
|-------|-------|------------------------|------------------------|----------------------------|----------------------------|----------------------------|
| C1    | 1400  | 338                    | 393                    | 7794                       | 6354                       | 22                         |
| C2    | 1400  | 328                    | 381                    | 7794                       | 6354                       | 22                         |
| C3    | 1400  | 317                    | 369                    | 7794                       | 6354                       | 22                         |
| C4    | 1400  | 340                    | 393                    | 7794                       | 7268                       | 23                         |
| C5    | 1400  | 330                    | 381                    | 7794                       | 7268                       | 23                         |
| C6    | 1400  | 319                    | 369                    | 7794                       | 7268                       | 23                         |
| C1    | 1600  | 0                      | 393                    | 7666                       | 5989                       | 8                          |
| C2    | 1600  | 0                      | 381                    | 7666                       | 5989                       | 8                          |
| C3    | 1600  | 0                      | 369                    | 7666                       | 5989                       | 8                          |
| C4    | 1600  | 260                    | 393                    | 7666                       | 6914                       | 11                         |
| C5    | 1600  | 252                    | 381                    | 7666                       | 6914                       | 11                         |
| C6    | 1600  | 244                    | 369                    | 7666                       | 6914                       | 11                         |

**Table S2:** Properties of the core models assumed to obtain the preferred lunar interior models discussed in the main text. Note that the outer core densities do not necessarily correlate with the outer core density. This is because the uncertainty in the outer core density is very high and extreme values were used to minimize/maximize bulk core density. The shown values are consistent with the assumed outer core S contents within the uncertainty of the values reported by Antonangeli et al. (2015).